\documentclass[unnumsec,webpdf,contemporary,large]{oup-authoring-template}

\usepackage[utf8]{inputenc}
\usepackage[T1]{fontenc}
\usepackage{soul}
\usepackage{booktabs}
\usepackage{hyperref}
\usepackage{cleveref}
\usepackage{enumerate}

\usepackage{titlesec}
\definecolor{jnlclr}{cmyk}{0,.94,.91,0.38}
\titleformat*{\section}{\secsize\bfseries\color{jnlclr}}
\titleformat*{\subsection}{\subsecsize\color{jnlclr}}
\hypersetup{
    frenchlinks=true,
    colorlinks=true,
    linkcolor=jnlclr,
    citecolor=jnlclr,
    urlcolor=jnlclr
}

\usepackage[final]{microtype}

\DeclareMathOperator{\sign}{sign}
\newcommand{\ang}{\textup{\AA}}
\newcommand\ddfrac[2]{\frac{\displaystyle #1}{\displaystyle #2}}

\graphicspath{{figures/}}

\begin{document}

\journaltitle{Microscopy and Microanalysis}
\DOI{}
\copyrightyear{2024}
\pubyear{2014}
\access{Advance Access Publication Date: Day Month Year}
\appnotes{Paper}

\firstpage{1}

\title[Iterative Phase Retrieval Algorithms for STEM]{Iterative Phase Retrieval Algorithms for Scanning Transmission Electron Microscopy}

\author[1,2,$\ast$,$\dagger$]{Georgios Varnavides\ORCID{0000-0001-8338-3323}}
\author[2,3,4,$\ast$]{Stephanie M. Ribet\ORCID{0000-0002-7117-066X}}
\author[5,6]{Stephen E. Zeltmann}
\author[7,8]{Yue Yu\ORCID{0000-0002-3248-9678}}
\author[2]{Benjamin H. Savitzky}
\author[2,5,9]{Dana O. Byrne}
\author[2,5,10]{Frances I. Allen}
\author[3,4,11]{Vinayak P. Dravid}
\author[2,5]{Mary C. Scott}
\author[2,$\ddagger$]{Colin Ophus\ORCID{0000-0003-2348-8558}}

\address[  1]{
\orgdiv{Miller Institute for Basic Research in Science}, \orgname{University of California}, \orgaddress{\street{Berkeley}, \postcode{94720}, \state{CA}, \country{USA}}
\\
}
\address[  2]{
\orgdiv{National Center for Electron Microscopy}, \orgname{Molecular Foundry, Berkeley Lab}, \orgaddress{\street{Berkeley}, \postcode{94720}, \state{CA}, \country{USA}}
\\
}
\address[  3]{
\orgdiv{Department of Materials Science and Engineering}, \orgname{Northwestern University}, \orgaddress{\street{Evanston}}, \postcode{60208}, \state{IL}, \country{USA}
\\
}
\address[  4]{
\orgdiv{International Institute of Nanotechnology}, \orgname{Northwestern University}, \orgaddress{\street{Evanston}, \postcode{60208}, \state{IL}, \country{USA}}
\\
}
\address[  5]{
\orgdiv{Department of Materials Science and Engineering}, \orgname{University of California, Berkeley}, \orgaddress{\street{Berkeley}, \postcode{94720}, \state{CA}, \country{USA}}
\\
}
\address[  6]{
\orgdiv{Platform for the Accelerated Realization, Analysis, and Discovery of Interface Materials}, \orgname{Cornell University}, \orgaddress{\street{Ithaca}, \postcode{14853}, \state{NY}, \country{USA}}
\\
}
\address[  7]{
\orgdiv{School of Applied and Engineering Physics}, \orgname{Cornell University}, \orgaddress{\street{Ithaca}, \postcode{14853}, \state{NY}, \country{USA}}
\\
}
\address[  8]{
\orgdiv{Chan Zuckerberg Institute for Advanced Biological Imaging}, \orgaddress{\street{Redwood City}, \postcode{94063}, \state{CA}, \country{USA}}
\\
}

\address[  9]{
\orgdiv{Department of Chemistry}, \orgname{University of California, Berkeley}, \orgaddress{\street{Berkeley}, \postcode{94720}, \state{CA}, \country{USA}}
\\
}

\address[ 10]{
\orgdiv{California Institute for Quantitative Biosciences}, \orgname{University of California,Berkeley} \orgaddress{\street{Berkeley}, \postcode{94720}, \state{CA}, \country{USA}}
\\
}

\address[ 11]{
\orgdiv{The NUANCE Center}, \orgname{Northwestern University}, \orgaddress{\street{Evanston}, \postcode{30208}, \state{IL}, \country{USA}}
\\
}
\corresp[$\ast$]{These authors contributed equally to this work.\\}
\corresp[$\dagger$]{\href{gvarnavides@berkeley.edu}{gvarnavides@berkeley.edu}\\}
\corresp[$\ddagger$]{\href{cophus@gmail.com}{cophus@gmail.com}\\}

\received{Date}{0}{Year}
\revised{Date}{0}{Year}
\accepted{Date}{0}{Year}

\abstract{
Scanning transmission electron microscopy (STEM) has been extensively used for imaging complex materials down to atomic resolution. 
The most commonly employed STEM modality, annular dark-field imaging, produces easily-interpretable contrast, but is dose-inefficient and produces little to no discernible contrast for light elements and weakly-scattering samples. 
An alternative is to use STEM phase retrieval imaging, enabled by high speed detectors able to record full images of a diffracted STEM probe over a grid of scan positions. 
Phase retrieval imaging in STEM is highly dose-efficient, enabling the measurement of the structure of beam-sensitive materials such as biological samples. 
Here, we comprehensively describe the theoretical background, algorithmic implementation details, and perform both simulated and experimental tests for three iterative phase retrieval STEM methods: focused-probe differential phase contrast, defocused-probe parallax imaging, and a generalized ptychographic gradient descent method implemented in two and three dimensions. 
We discuss the strengths and weaknesses of each of these approaches by comparing the transfer of information using analytical expressions and numerical results for a white-noise model. 
This presentation of STEM phase retrieval methods aims to make these methods more approachable, reproducible, and more readily adoptable for many classes of samples.
}


\maketitle

\section{Introduction}

The ``phase problem'' -- describing the loss of phase information of complex valued scattering samples during intensity measurements -- is a well-known challenge in many imaging and diffraction fields including crystallography, astronomy, and microscopy~\citep{fienup1982phase}.
\textit{Phase contrast imaging} techniques, such as Zernike phase contrast microscopy (PCM)~\citep{zernike1942phase, danev2001transmission} and differential phase contrast (DPC)~\citep{dekkers1974differential}, attempt to solve this by converting phase variations in the object plane into intensity variations in the imaging plane.
Conversely, \textit{phase retrieval} techniques attempt to recover the missing phase information by leveraging redundant information in the dataset and prior information such as finite support and positivity in the form of projections and regularization constraints~\citep{fienup1982phase,Rodenburg2019}.
Such techniques often involve iterative reconstruction algorithms and are thus are also referred to as \textit{computational} or \textit{lensless} imaging techniques ~\citep{boominathan2022recent}.
Phase retrieval techniques offer particular promise in scanning transmission electron microscopy (STEM), because they enable the observation of otherwise imperceptible signals from weakly scattering and dose sensitive samples~\citep{pennycook2019,Zhou2020,scheid2023,ophus2023quantitative}.

\begin{figure*}
\includegraphics[width=\textwidth]{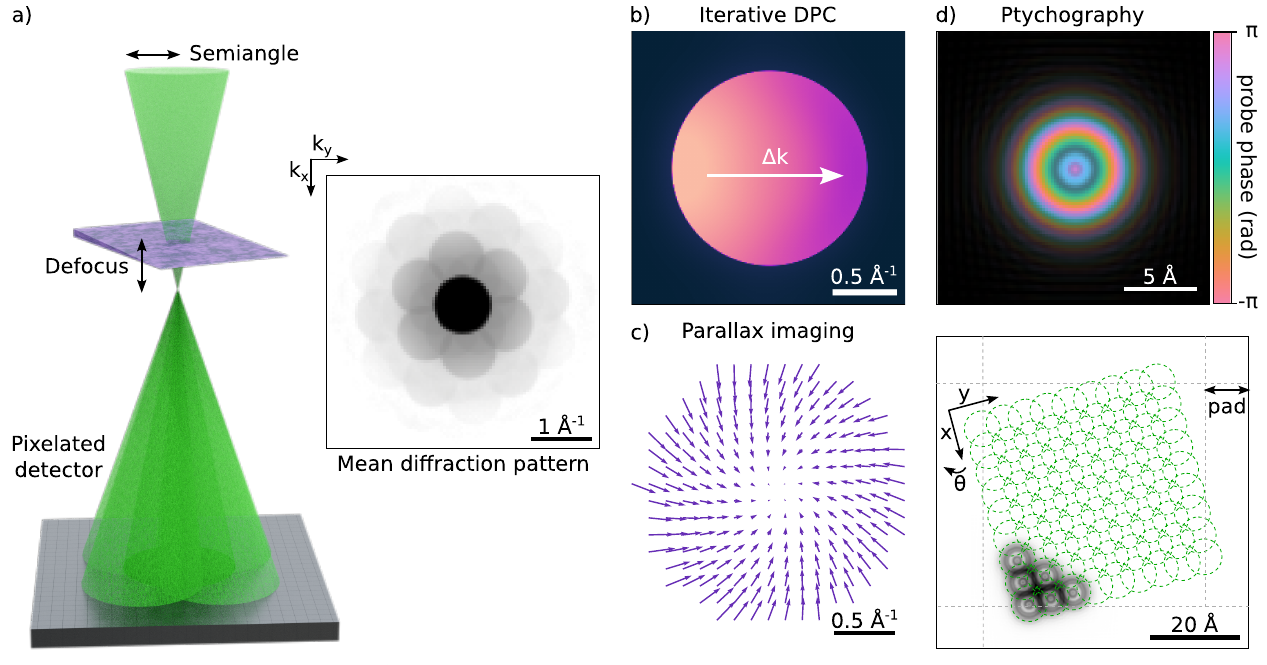}
\caption{a) 4D-STEM geometry.
A converged electron probe is scanned across a sample, with diffraction intensities being recorded on a pixelated detector at the far-field.
b) Beam intensity center of mass shift used in DPC, resulting from long-range electrostatic fields.
c) Cross-correlation vector field for virtual bright-field alignment used during parallax reconstruction.
d) Probe and object field-of-view parameters relevant for ptychography.
}
\label{fig:overview}
\end{figure*}

While early phase retrieval microscopy techniques such as coherent diffractive imaging (CDI) used parallel-beam illumination and a single diffraction measurement~\citep{miao1999extending}, significant advantages can be conferred by the diversity of information obtained in scanning a converged beam across the sample and collecting diffraction intensities at each probe position.
In electron microscopy this imaging mode is often referred to as 4D scanning transmission electron microscopy (4D-STEM) (\cref{fig:overview}a), due to the dimensionality of the resulting dataset, having two (real-space) scan dimensions and two (reciprocal-space) diffraction dimensions~\citep{ophus2019four}.

The rapid development of phase retrieval in electron microscopy in the last decade is due in part to the fact that many of the guiding principles translate well across similar techniques developed for longer-wavelength illumination such as x-ray and light microscopy, and more broadly from the field of non-convex optimization~\citep{rodenburg2004,thibault2008high,boyd2014}. 
However, various nuances of the scattering physics, 4D-STEM geometry, and electron optics suggest that algorithms adapted from other fields can benefit from domain-specific development, inclduing efficient regularization constraints.

Here, we present a self-consistent framework for various phase-retrieval algorithm implementations designed for 4D-STEM datasets, with a particular emphasis on robustness against common experimental artifacts, available in the open-source 4D-STEM analysis software package \texttt{py4DSTEM}~\citep{savitzky2021py4dstem}.
In particular, we highlight three classes of iterative phase-retrieval techniques: i) iterative DPC (\cref{fig:overview}b), ii) parallax imaging (\cref{fig:overview}c), and iii) a suite of ptychographic reconstructions in two- and three-dimensions (\cref{fig:overview}d).
The implementations discussed below include a combination of previously published algorithms as well as original research, and we highlight our contributions to the field throughout the text.

The manuscript is structured as follows:
First, in~\cref{sec:iterative-dpc,sec:parallax} we introduce the theory for the iterative DPC and parallax phase-retrieval techniques, and demonstrate their utility on simulated datasets for focused-probe and defocused-probe geometries, respectively.
In addition to the reconstructed phases, these techniques provide valuable tools to estimate necessary preprocessing parameters for more involved ptychographic reconstructions, namely the relative rotation and alignment of the diffraction plane coordinate system with respect to the scan coordinate system and low-order aberrations such as defocus.
We then introduce the theory for single-slice ptychography in~\cref{sec:single-slice_ptycho} and discuss common preprocessing parameters and their impact on the resulting reconstruction.
Next, we compare the information transfer of the three phase retrieval techniques by evaluating their contrast transfer functions in the presence of random and systematic errors, using a numerical model of a white noise phase object in~\cref{sec:ctf}.
Finally, we demonstrate the performance of our phase retrieval implementations against experimental datasets on materials science and biological samples in~\cref{sec:experimental-results}.
Various algorithmic and experimental considerations, such as domain-specific regularization constraints for the illuminating probe and reconstructed object in ptychographic reconstructions, are discussed in appendix~\cref{sec:algorithmic-considerations,sec:experimental-considerations}.
We also include discussions on how the assumptions of single-slice ptychography can be relaxed to include partial-coherence (appendix~\cref{sec:mixed-state_ptycho}), depth-sectioning (appendix~\cref{sec:multi-slice_ptycho}), and three-dimensional scattering (appendix~\cref{sec:overlap-tomo}).

\section{Iterative Differential Phase Contrast} \label{sec:iterative-dpc}

\begin{figure*}
    \centering
    \includegraphics[width=\textwidth]{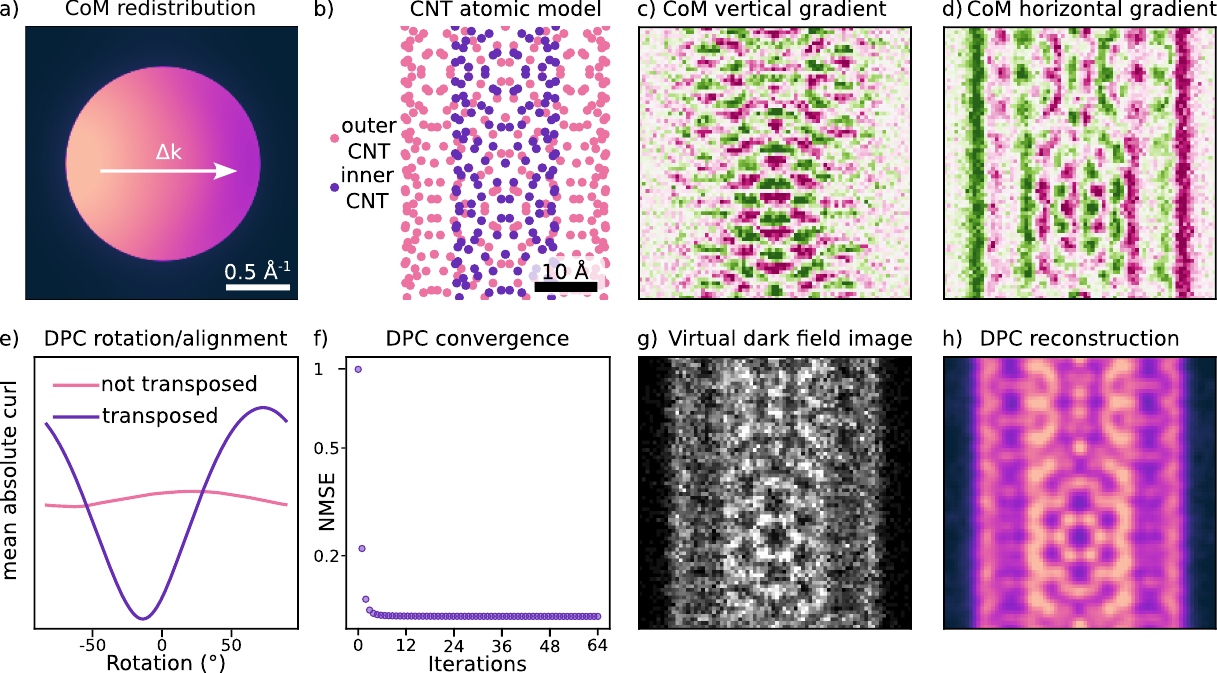}
    \caption{a) Schematic showing center of mass shifts when probe interacts with sample.
    b) Atomic model of double walled carbon nanotube.
    Virtual CoM images along c) vertical- and d) horizontal-directions with green and purple indicating +ve and -ve shifts respectively.
    e) Mean CoM curl as a function of real/reciprocal-space rotation alignment.
    f) Convergence of iterative DPC algorithm.
    g) Virtual dark-field image and h) iterative DPC reconstruction.
    }
    \label{fig:dpc}
\end{figure*}

When an electron beam interacts with a sample's electrostatic potential the center of mass (CoM) of the momentum distribution of the beam intensity shifts.
The nature of the CoM shift is related to the probe and feature size (\cref{fig:dpc}a)~\citep{cao2018theory}, and can be used to reconstruct the sample's electrostatic potential.  
Atomic-scale features tend to lead to redistribution of signal within the bright-field disk, while long-range (slowly-varying) fields cause near-rigid shifts of the entire bright-field disk.
Early techniques captured this change in center of mass through segmented detectors, using a technique called differential phase contrast (DPC)~\citep{dekkers1974differential}. 
Further developments in hardware have led to the design of more complex segmented detectors with multiple annular rings, which provide radial in addition to angular sensitivity~\citep{shibata2010new, seki2021toward}. 

Soon after the development of segmented detector DPC, it was recognized that integrating the signal from a pixelated detector with a center of mass response function can provide an alternative route to capturing the phase of the beam~\citep{waddell1979linear}. 
While such a detector was not physically realizable at the time, today they are widely available for 4D-STEM experiments. 
Reconstructing the phase-shift of thin specimens using a virtual 4D-STEM detector as described below goes by various names including first moment STEM ~\citep{muller2019comparison}, integrated center of mass (ICoM)~\citep{li2022integrated}, and even simply differential phase contrast (DPC)~\citep{yang2015efficient, savitzky2021py4dstem}.
Despite the fact that the technique is a diffractive imaging technique which does not modulate contrast in the imaging plane, in what follows we will refer to it as \textit{iterative DPC} for historical reasons and to make the connection with segmented detectors clearer.

\subsection{Inverse Problem Formalism} \label{subsec:inverse-problem-dpc}
 
We start by considering a 4D dataset composed of diffraction intensities $I(\boldsymbol{R},\boldsymbol{k})$, where  $\boldsymbol{R}$ and  $\boldsymbol{k}$ denote the real-space probe position and reciprocal-space spatial frequency, respectively.
We form a vectorial virtual image using a first moment detector $\boldsymbol{k}$:

\begin{align}
 \mathcal{I}_{\alpha}^V(\boldsymbol{R}) = 
    \int k_{\alpha} \, I(\boldsymbol{R},\boldsymbol{k}) \,  d\boldsymbol{k}, \label{eq:dpc-virtual}
\end{align}
where $\alpha$ denotes each of the two Cartesian directions $x$ and $y$ and the virtual CoM image, $\mathcal{I}_{\alpha}^V(\boldsymbol{R})$, is proportional to the gradient of the electrostatic sample potential $\partial_{\alpha} V(\boldsymbol{R})$.
The electrostatic sample potential can be reconstructed by Fourier-integrating~\cref{eq:dpc-virtual} iteratively~\citep{frankot1988method}:

\begin{align}
     V'(\boldsymbol{R}) = V(\boldsymbol{R}) + \beta \Re\left\{\mathcal{F}^{-1}
     \left[\frac{\mathrm{i} \boldsymbol{k}}{k^2} \cdot \mathcal{F}\left[ \boldsymbol{\mathcal{I}}^V(\boldsymbol{R})-\nabla V(\boldsymbol{R})\right] \right]\right\}, \label{eq:dpc-fourier-integration} 
\end{align}
where $\mathcal{F}$ is the two-dimensional Fourier transform operator, $\nabla$ is the gradient operator which can be computed numerically using a centered finite-difference, $\beta$ is the step size, $k$ is the spatial frequency magnitude, $\Re\{\cdot\}$ denotes the real-part of the expression and~$'$ indicates the next iteration.
Note that here, and in what follows, we express the phase-shift imparted by projected potentials in units of radians.
We define a self-consistent error metric using~\cref{eq:dpc-fourier-integration}, by comparing the current estimate of the finite-difference gradient against the experimental CoM:

\begin{align}
     \mathcal{E} = \ddfrac{\sum_{\boldsymbol{R}} \sum_{\alpha} \left| \partial_{\alpha} V(\boldsymbol{R})- \mathcal{I}^V_{\alpha}(\boldsymbol{R})\right|^2}{\sum_{\boldsymbol{R}} \sum_{\alpha} \left|\mathcal{I}^V_{\alpha}(\boldsymbol{R})\right|^2}.
\end{align}

To avoid periodic artifacts arising from the use of Fourier transforms, we pad and mask the reconstructed object at every iteration~\citep{savitzky2021py4dstem}.
Finally, we implement a simple backtracking algorithm where the step-size is automatically halved if the error increases relative to the previous iteration and the current update step is rejected.

\subsection{Preprocessing and Results} \label{subsec:dpc-preprocessing-results}

Due to the helical path accelerated electrons take along the optic axis, as well as the orientation of the detector in the STEM column relative to the beam raster grid, there is typically a rotational offset between the real- and reciprocal-space coordinate systems.
Since the gradient of the electrostatic potential is expected to be a conservative vector field, we can solve for this rotation by minimizing the curl or maximizing the divergence of the vectorial CoM virtual image as a function of rotational offset~\citep{savitzky2021py4dstem, zachman2022robust}. 
These techniques have a $180^\circ$ ambiguity in the relative rotation which can be solved by requiring the phase-shift to be positive everywhere except over vacuum (approximately true for most samples), or by performing an independent ``shadow-image'' calibration using a defosued ronchigram~\citep{savitzky2021py4dstem}.  
While solving for the relative rotation angle is only somewhat important for iterative DPC, ptychographic reconstructions are especially sensitive to this value (see~\cref{sec:single-slice_ptycho}).

\Cref{fig:dpc} illustrates our iterative DPC approach with a simulated double-walled carbon nanotube (CNT) structure (\cref{fig:dpc}b). 
We introduce a -17$^\circ$ rotational offset between real-space and reciprocal-space and transpose the two axes of the diffraction intensities. 
This transpose is common to many experimental set-ups depending on the detector readout and STEM scanning direction, and can be solved for using with the same curl minimization approach. 
The rotationally-aligned and transposed CoM images are shown in~\cref{fig:dpc}c-d. 
The curl minimization algorithm recovers the proper alignment (\cref{fig:dpc}e), as seen by the CoM images' contrast being aligned with the $\hat{x}$- and $\hat{y}$-axes respectively. 
The convergence profile of our DPC reconstruction is shown in~\cref{fig:dpc}f, illustrating a large drop off in error after the first few iterations. 
Comparing the virtual dark-field image in~\cref{fig:dpc}g and the iterative DPC reconstruction in~\cref{fig:dpc}h from the same dataset, we can observe how much more information-rich the iterative DPC reconstruction is for this simulated dataset. 

There are several advantages to using iterative DPC for recovering sample-induced phase-shifts. 
First, compared to other phase-retrieval techniques it is relatively straightforward and computationally-inexpensive, so it can be used to quickly characterize the phase-shift imparted by the sample. 
Moreover, similar to other phase-contrast techniques, it is dose-efficient and enables linear contrast (see~\cref{sec:ctf}), although additional considerations are required for quantitative analysis. 
However, iterative DPC reconstructions only solve for the phase of the sample and cannot deconvolve the probe's wavefunction. 
Therefore, any aberrations in the probe, including defocus, will affect the transfer of information in the reconstruction. 
Lastly, the pixel-size of the reconstruction is limited by the real-space step size of the STEM probe and the absolute resolution limit is set by twice the convergence semi-angle (see~\cref{sec:ctf}).

\begin{figure*}
\includegraphics[width=\textwidth]{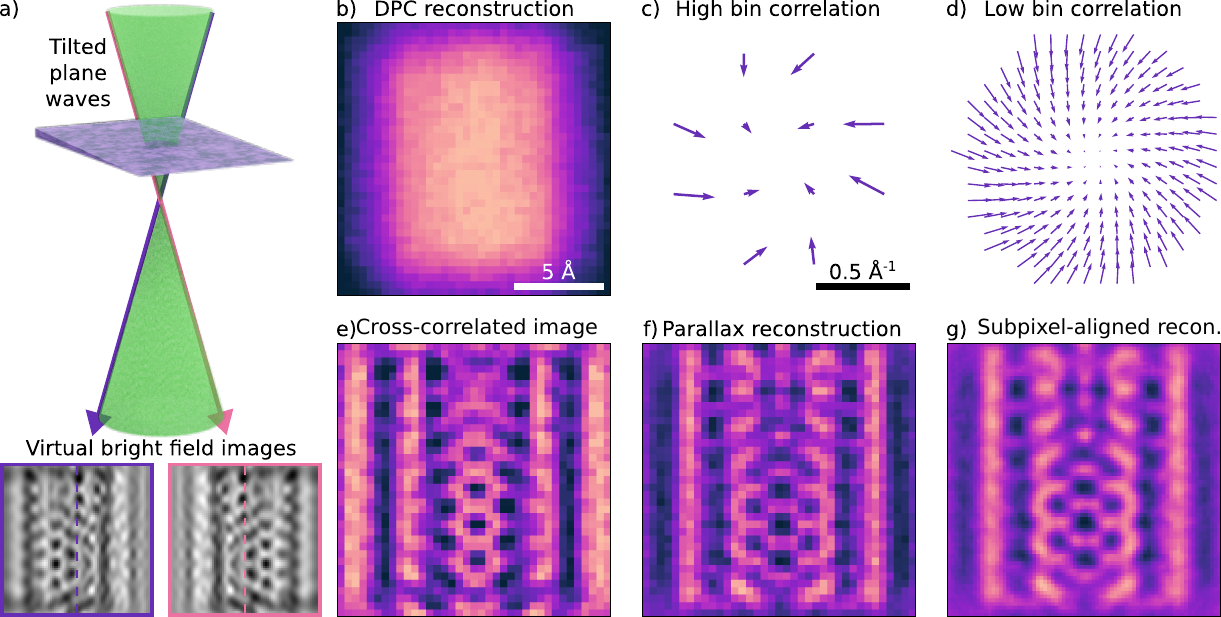}
\caption{
(a) Parallax schematic.
Virtual images from individual positions on the central disk represent incoming plane waves at different angles, which appear as specimen shift/tilt.
Vertical dashed lines added at field of view midpoint as guides to the eye.
(b) Iterative DPC reconstruction. 
(c-d) Vector fields for cross-correlating images at different levels of binning. 
(e) Aligned cross-correlated bright-field image and (f) CTF-corrected image showing strong contrast from the CNT. 
(g) Upsampled parallax reconstruction highlights the higher resolution. 
}
\label{fig:parallax}
\end{figure*}

\section{Parallax Imaging} \label{sec:parallax}

Despite the benefits described in~\cref{sec:iterative-dpc} of using iterative DPC for phase retrieval in 4D-STEM datasets, the reconstruction quality is limited by the aberrations in the incident electron probe. 
\Cref{fig:parallax}b illustrates an iterative DPC reconstruction for the same CNT structure with a defocused probe. 
Unlike in~\cref{fig:dpc}, it is not possible to resolve atomic features, despite using similar sampling. 
As we will see in~\cref{sec:single-slice_ptycho}, deliberately defocusing the probe can be a way of relaxing real-space sampling requirements. 

In this section, we present another phase retrieval technique, we term \textit{parallax imaging}, which can be thought of as a weak-phase approximation to the more general \textit{scattering matrix reconstruction} methods~\citep{brown2018structure, pelz2021phase, findlay2021scattering}.
This approach is also referred to as \textit{tilt-corrected bright-field STEM} (tc-BF STEM) and has been demonstrated as an effective imaging technique for biological samples~\citep{spoth2017dose, yu2022dose, yu2024dose}.
Parallax imaging can estimate the aberration coefficients for highly-aberrated probes, and in doing so more faithfully reconstruct the phase imparted by the sample~\citep{ophus2019advanced}. 
Good estimates of these low-order aberrations are often useful or even necessary for high-quality ptychographic reconstructions and can be hard to obtain during data acquisition.

Each pixel in the diffracted bright-field disk can be thought of as a plane wave of electrons which impacted the sample at a slightly different angle, up to a maximum tilt given by the convergence semiangle.
Geometrically, this corresponds to illuminating different positions on the sample surface, with increasing stride per pixel for increasing defocus, and thus named after the ``stellar parallax'' phenomenon~\citep{bessel1838bestimmung, siebert2005early}.
\Cref{fig:parallax}a illustrates the effect for the simulated CNT sample shown in~\cref{fig:dpc}b.
Two virtual images of the CNT created from different bright-field pixels are shown, coloured purple and pink respectively, highlighting the relative shift and tilt apparent in the CNT structure. 
Note that while the virtual images are shown at infinite dose for visual clarity, the algorithm is robust to low dose conditions as we demonstrate below. 

\subsection{Inverse Problem Formalism} \label{subsec:inverse-problem-parallax}

We start by forming a virtual image from each pixel within the bright-field disk.
Each of these virtual images will be offset from the optical axis and can be computationally aligned to the center pixel using the geometric parallax operator. 
In short, the operator corrects the geometric intercepts of each virtual image to be properly focused by aligning the virtual images. 
In our iterative algorithm, we typically start with a heavily-binned dataset to enable alignment of low-dose datasets and progressively decrease the amount of binning, using the previous iteration's shifts as a starting point for cross-correlating the virtual images. 
\Cref{fig:parallax}c shows the relative shifts of the virtual images with a high degree of binning, and as we iteratively decrease the level of binning, we recover a finely-sampled vector field used to align the simulated CNT dataset (\cref{fig:parallax}d).
The high signal-to-noise aligned bright-field image is show in \cref{fig:parallax}e.

The cross-correlation vector shifts $\boldsymbol{w}(\boldsymbol{k})$ can be expressed using the Cartesian derivatives of the aberration surface 

\begin{align}
    \boldsymbol{w}(\boldsymbol{k}) = \nabla \chi(\boldsymbol{k}) \label{eq:chi-derivatives},
\end{align}
where $\chi(\boldsymbol{k})$ can be written as a function of spatial frequency:

\begin{align}
    \chi(\boldsymbol{k}) = \frac{2\pi}{\lambda} \sum_{m,n} \frac{\left(\lambda \; k \right)^{m+1}}{m+1} \Big( & C_{m,n}^x \cos{\left[n \times \mathrm{atan2}\left(k_y, k_x\right) \right]}  + \Big. \notag \\ 
    \Big. & C_{m,n}^y \sin{\left[n \times \mathrm{atan2}\left(k_y, k_x\right) \right]}\Big) \label{eq:chi-expansion}.
\end{align}

Here,  $\lambda$ is the electron wavelength, $C_{m,n}^{x/y}$ are the coefficients of the two orthogonal aberrations of radial order $m+1$ and angular order $n$ in units of \ang ngstr\"{o}ms, and $\mathrm{atan2}\left(k_y, k_x\right)$ is the arctangent function which returns the correct sign of the aberration axis in all quadrants.
When $n=0$ the aberrations are radially symmetric (e.g. defocus, spherical aberration, etc.) and no $C_{m,0}^y$ term is necessary.
\Cref{eq:chi-derivatives,eq:chi-expansion} form a linear system of equations, which can be used to solve for the aberration coefficients~\citep{cowley1979coherent,lupini2010aberration,lupini2016}.

It is instructive to investigate the aberration coefficients fit explicitly for the lowest-order aberrations, namely defocus and astigmatism ($m_{\mathrm{max}}=1$).
Let the initial pixel positions, $\boldsymbol{v}(\boldsymbol{k})$, and measured vector shifts, $\boldsymbol{w}(\boldsymbol{k})$, be denoted by the real-valued arrays $V \in \mathbb{R}^{G \times 2}$ and $W \in \mathbb{R}^{G \times 2}$ respectively, where $G$ is the number of bright-field pixels.
The linear system of equations can then be expressed using the operator $H \in \mathbb{R}^{2 \times 2}$:

\begin{align}
    V H = W,
\end{align}
which can be estimated using least-squares and further decomposed into rotational $U$ and radial components $P$:

\begin{align}
    \hat{H} &= \left(V^TV\right)^{-1}V^T W \label{eq:least-squares} \\
     &= UP. \label{eq:polar_decomposition}
\end{align}

These can further be used to recover estimates for the $\theta$ (see~\cref{sec:iterative-dpc}), defocus $C_{1,0}$, and astigmatism $C_{1,2}^{x/y}$ coefficients:

\begin{align}
    P & = \begin{bmatrix}
        C'_1 & A'_1 \\
        A''_1 & C''_1 
        \end{bmatrix},
    \quad
    U = \begin{bmatrix}
        cos(\theta) & sin(\theta) \\
        -sin(\theta) & cos(\theta) 
        \end{bmatrix}, \\
    C_{1,0} &= \frac{C'_1 + C''_1}{2}, \quad C_{1,2}^x = \frac{A'_1 - A''_1}{2}, \quad\text{and}\; 
    C_{1,2}^y = \frac{A'_1 + A''_1}{2}. \notag
\end{align}

\subsection{Preprocessing and Results} \label{subsec:parallax-preprocessing-results}

We can use our aberration coefficient estimates to apply a contrast transfer function (CTF) correction to the reconstruction (\cref{fig:parallax}f) using the following phase-flipping filter in Fourier space~\citep{lupini2016}: 

\begin{align}
    \xi(\boldsymbol{k}) &=  \sign \left[ \sin\left(\chi^{\mathrm{even}}(\boldsymbol{k})\right) \right] \times \exp \left[-\mathrm{i} \chi^{\mathrm{odd}}(\boldsymbol{k}) \right],
\end{align}
where $\chi^{\mathrm{even}}(\boldsymbol{k})$/$\chi^{\mathrm{odd}}(\boldsymbol{k})$ are even/odd components of~\cref{eq:chi-expansion} and $\sign\left[\cdot\right]$ returns the sign of its argument.
The complex-valued phase ramp imparted by the odd contribution ensures the filter remains conjugate symmetric, returning a real-valued signal.

Finally, we note that the virtual images can be upsampled after alignment using kernel-density estimation of the cross-correlation vector field~\citep{ophus2016}, creating a high signal-to-noise reconstruction (\cref{fig:parallax}g). 
The final upsampled reconstruction in~\cref{fig:parallax}g illustrates the structure of the simulated CNT dataset. 
This is a dose-efficient technique since it uses all the scattered information inside the bright-field disk.

More complex deconvolution of signal in the bright-field disk is possible through scattering matrix reconstruction, which has been used for 3D-reconstructions of thicker samples~\citep{brown2022three}. 
Similarly, a larger convergence angle will improve the depth-of-field, and the parallax operator can be applied to the upsampled reconstructions to study the out of plane structure of materials \citep{terzoudis2023resolution}.

Overall, parallax imaging is a dose-efficient and relatively computationally-inexpensive approach for recovering the structure of weak-phase specimens in a defocused dataset.
Unlike iterative DPC, the resolution of the upsampled reconstruction is not limited by the scan step-size, but instead is equal to twice the convergence semi-angle similarly to other phase contrast STEM methods~\citep{ophus2016efficient}. 
For low-dose experiments with weakly scattering samples, parallax additionally provides a good estimate for low-order aberrations, which may be used as input for subsequent ptychographic reconstructions. 
Due to the low-computational overhead of parallax reconstructions, they can also be incredibly beneficial for checking microscope alignments and sample conditions during data acquisition.

\section{Single-slice Ptychography} \label{sec:single-slice_ptycho}

Electron ptychography was proposed by Walter Hoppe in 1969~\citep{Hoppe1969a,Hoppe1969c,Hoppe1969b}, with the first STEM proof-of-principle some 25 years later demonstrating the advantages of the method over other phase retrieval techniques in achieving resolution beyond the `information limit'~\citep{Nellist1995}.
These early methods developed into two non-iterative algorithms still in-use today, namely \textit{single side-band} (SSB)~\citep{pennycook2015efficient} and \textit{Wigner-distribution deconvolution} (WDD)~\citep{rodenburg1992}.
While these techniques enable dose-efficient imaging at high resolution, they require stringent sampling conditions in both real- and reciprocal-space.
In what follows, we instead focus on iterative ptychographic techniques which relax these requirements.

Compared to the phase-retrieval algorithms introduced so far, the iterative ptychographic algorithms described below offer two distinct advantages.
First, since both the sample transmission function and the incoming- wave illumination are jointly reconstructed, the algorithms can overcome the resolution limitations set by residual aberrations.
In-fact, probes are often deliberately defocused during ptychographic data acquisitions to increase probe overlap and thus information redundancy (see~\cref{subsec:probe-overlap-dose}), enabling dose fractionation and faster acquisition speeds.
Secondly, since the reconstruction resolution is set by the largest acquired spatial frequency in the diffraction intensities (see~\cref{subsec:ptycho-preprocessing}), this enables super-resolution imaging.
Despite these advantages, iterative ptychographic reconstructions require more complex algorithms and are significantly more computationally-expensive, and the reconstruction can fail if the dataset does not contain sufficient redundancy~\citep{chen2020mixed}.

\begin{figure*}
\includegraphics[width=\textwidth]{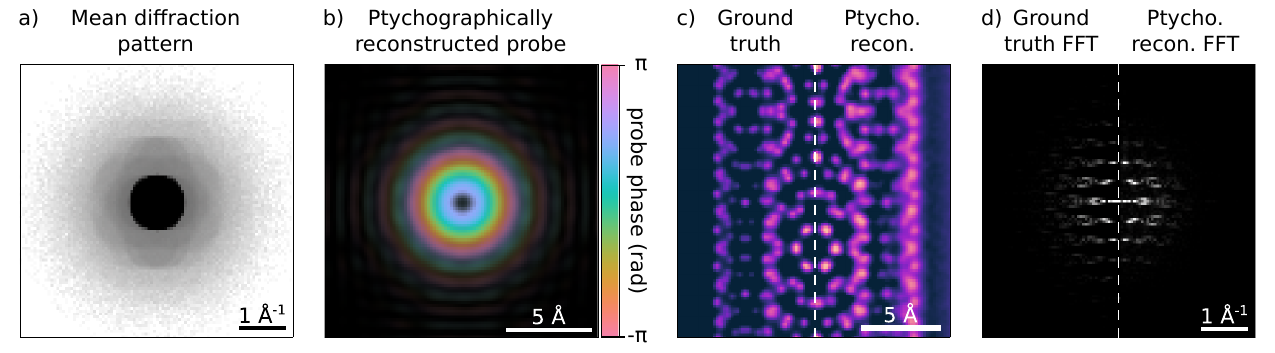}
\caption{Single-slice ptychographic reconstruction of simulated double-walled carbon nanotube showing a) the mean diffraction pattern using finite dose of $10^6$e$^-$/\AA$^2$,
b) the reconstructed probe using a $150 \ang$ initial defocus,
c) the reconstructed object and d) spatial frequencies.
In c) and d) we compare the ground-truth potential (left) and the reconstructed object (right).
}
\label{fig:single-slice}
\end{figure*}

\subsection{Inverse Problem Formalism} \label{subsec:inverse-problem-ptycho}

In this section, we set up the inverse problem for the simplest ptychographic formulation and adjust the models as appropriate in the following sections to capture more complex scattering physics.
While as we saw in~\cref{sec:iterative-dpc,sec:parallax}, typical 4D-STEM experimental diffraction intensities are acquired from a 2D raster scan, this is not strictly required and complicates ptychographic notation.
Instead, in the following sections we consider a dataset, $I(\boldsymbol{R}_m,\boldsymbol{k}) = I_m(\boldsymbol{k})$, of diffraction intensity measurements collected from any $M$ probe positions, $\boldsymbol{R}_m$.
Single-slice ptychography assumes the exit-wave, $\psi_m(\boldsymbol{r})$, is given by a simple multiplication of the complex-valued incoming-wave and the complex-valued sample transmission function, henceforth referred to as the probe, $\mathcal{P}(\boldsymbol{r})$, and object, $\mathcal{O}(\boldsymbol{r}-\boldsymbol{R}_m)$, respectively, propagated to the farfield detector:

\begin{align}
    \psi_m(\boldsymbol{r}) &= \Pi_O\left[\mathcal{P},\mathcal{O}\right] = \mathcal{P}(\boldsymbol{r})\mathcal{O}(\boldsymbol{r}-\boldsymbol{R}_m) \label{eq:overlap-projection} \\
    J_m(\boldsymbol{k}) &= \left| \mathcal{F}\left[\psi_m(\boldsymbol{r})\right] \right|^2, \label{eq:model-intensity}
\end{align}
where $\Pi_O$ is referred to as the "overlap projection" and $J_m(\boldsymbol{k})$ is the estimated beam intensity at probe position $\boldsymbol{R}_m$.
The inverse problem is to find probe and object operators which minimize the distance between the modelled and measured diffraction intensities given by an appropriate error metric such as:

\begin{align}
    \mathcal{E} = \ddfrac{\sum_{m,\boldsymbol{k}}\left|\sqrt{I_m(\boldsymbol{k})} - \sqrt{J_m(\boldsymbol{k})}\right|^2}{\sum_{m,\boldsymbol{k}} I_m(\boldsymbol{k})}. \label{eq:nmse-error}
\end{align}

From the many existing algorithms to solve this inverse problem~\citep{fienup1982phase,rodenburg2004,thibault2008high,maiden2009improved,maiden2017further}, we have implemented stochastic gradient descent and proximal gradient solvers due to their robustness and intuitive geometric interpretation, respectively.
The first step is recognizing that the Euclidean projection of the model estimate onto the measured data, i.e. the minimal modification to the current estimate of the complex-valued exit-wave to match the measured intensities, is given by replacing its Fourier amplitude with the square-root of the measured intensity while retaining its phase~\citep{fienup1982phase}:

\begin{align}
    \Pi_F\left [ \psi_m(\boldsymbol{r})\right] = \mathcal{F}^{-1}\left[\sqrt{I_m(\boldsymbol{k})} \mathcal{F}[\psi_m(\boldsymbol{r})] / \sqrt{J_m(\boldsymbol{k})}\right], \label{eq:fourier-projection}
\end{align}
where $\Pi_F$ is referred to as the ``Fourier projection.''
This allows us to define the appropriate inverse operators for the stochastic gradient descent and proximal gradient methods respectively:

\begin{align}
    \intertext{(Stochastic Gradient Descent)}
    \Delta \psi_m^{\mathrm{GD}}(\boldsymbol{r}) &= \Pi_F\left[\psi_m(\boldsymbol{r})\right]-\psi_m(\boldsymbol{r}) \label{eq:gd-psi} \\
    \mathcal{O}'(\boldsymbol{r}) &= \mathcal{O}(\boldsymbol{r}) + \ddfrac{\beta \sum_{m \in \mathcal{M}} \mathcal{P}^*(\boldsymbol{r})\Delta \psi_m^{\mathrm{GD}}(\boldsymbol{r})}{|| \mathcal{P}(\boldsymbol{r})||_{\alpha}} \label{eq:gd-update-o} \\
    \mathcal{P}'(\boldsymbol{r}) &= \mathcal{P}(\boldsymbol{r}) +\ddfrac{\beta \sum_{m \in \mathcal{M}} \mathcal{O}^*(\boldsymbol{r}-\boldsymbol{R}_m)\Delta \psi_m^{\mathrm{GD}}(\boldsymbol{r})}{|| \mathcal{O}(\boldsymbol{r}-\boldsymbol{R}_m)||_{\alpha}}, \label{eq:gd-update-p}
    \intertext{(Proximal Gradient)}
    \Delta \psi_m^{\mathrm{PG}}(\boldsymbol{r}) &= (1-a-b)\Delta \psi_m^{\mathrm{PG}}(\boldsymbol{r})+ a \psi_m(\boldsymbol{r}) \notag \\
    &\qquad \qquad + b\Pi_F\left[c \psi_m(\boldsymbol{r}) + (1-c)\Delta \psi_m^{\mathrm{PG}}(\boldsymbol{r})\right] \label{eq:proj-psi} \\
    \mathcal{O}'(\boldsymbol{r}) &= \ddfrac{\sum_m \mathcal{P}^*(\boldsymbol{r})\Delta \psi_m^{\mathrm{PG}}(\boldsymbol{r})}{|| \mathcal{P}(\boldsymbol{r})||_{\alpha}} \label{eq:proj-update-o} \\
    \mathcal{P}'(\boldsymbol{r}) &= \ddfrac{\sum_m \mathcal{O}^*(\boldsymbol{r}-\boldsymbol{R}_m)\Delta \psi_m^{\mathrm{PG}}(\boldsymbol{r})}{|| \mathcal{O}(\boldsymbol{r}-\boldsymbol{R}_m)||_{\alpha}} , \label{eq:proj-update-p}
    \intertext{where $\Delta \psi_m^{\mathrm{GD/PG}}$ are the gradients of the stochastic gradient descent and proximal gradient methods respectively, $\beta$ is the gradient-descent step-size, $\mathcal{M}$ denotes a batch of probe positions, $\left(a,b,c\right)$ parameterize a family of proximal gradient algorithms, and $\alpha$ is a scalar between 0-1 parametrizing the weight of the maximum probe overlap intensity in the $|| \cdot||_{\alpha}$ normalization:}
    || \cdot||_{\alpha} &= (1-\alpha)\sum_{m (\in \mathcal{M})} |\cdot|^2 + \alpha \sum_{m (\in \mathcal{M})} |\cdot|_{\mathrm{max}}^2 \label{eq:alpha-normalization}.
\end{align}

\subsection{Preprocessing and Results} \label{subsec:ptycho-preprocessing}

In  order to efficiently solve the system of~\cref{eq:gd-psi,eq:gd-update-o,eq:gd-update-p,eq:proj-psi,eq:proj-update-o,eq:proj-update-p,eq:alpha-normalization} the object, probe, and exit-wave functions introduced in~\cref{subsec:inverse-problem-ptycho} need to be stored numerically as evenly sampled arrays.
This introduces constraints during data acquisition and reconstruction which we outline below and ultimately control the numerical accuracy, computational cost, and resolution.

The acquired diffraction intensities are stored in the real-valued array, $I \in \mathbb{R}^{M \times Q \times Q}$, for an array of probe positions, $R \in \mathbb{R}^{M \times 2}$.
Note that when applying a rotation/transpose to align real and reciprocal space coordinate systems, we transform the probe positions and not the diffraction intensities, in order to avoid resampling the measured data (see~\cref{sec:iterative-dpc,sec:parallax}).

The reciprocal-space pixel size, $\Delta k$, defines the real-space extent, $e = 1/\Delta k$, of the complex-valued probe array, $P \in \mathbb{C}^{Q \times Q}$, where $\Delta k$ has to be selected such that $e$ is large enough to comfortably fit the probe without wrap-around artifacts (see~\cref{subsec:grdding}).
The largest acquired spatial frequency, $2 k_{\mathrm{max}} = \Delta k * Q$, defines the reconstruction real-space pixel size, $\Delta x = 1/(2 k_{\mathrm{max}})$, which together with the rotated/transposed probe positions define the object field-of-view in pixels.

The probe can be initialized with an estimate of the aberration coefficients and probe-forming aperture, obtained using microscope metadata for the probe semiangle or, better, from direct measurement of the probe under vacuum. 
The object array is then padded by as much as half the probe extent to define the size of the complex-valued array, $O \in \mathbb{C}^{S \times T}$, such that $S,T \geq Q$ (see~\cref{fig:overview}d).

In evaluating expressions of the form $\psi_m (\boldsymbol{r}) = \mathcal{P}(\boldsymbol{r}) \mathcal{O}(\boldsymbol{r}-\boldsymbol{R}_m)$, $Q \times Q$ region-of-interest patches are extracted from the larger $S \times T$ object array centered around $\boldsymbol{R}_m$, rounded to the nearest pixel, multiplied by the probe array, Fourier-shifted by any remaining sub-pixel offsets, and stored in the complex-valued exit-waves array, $\psi \in \mathbb{C}^{M \times Q \times Q}$.
Finally, note that the diffraction intensities, probe, and exit-waves are all internally stored as top-left corner-centered arrays to enable the unambiguous centering of even and odd dimension datasets and compatibility with fast Fourier transform libraries.

Consider the same double-walled CNT dataset reconstructed in~\cref{sec:iterative-dpc,sec:parallax}, simulated using $150 \ang$ defocus and a $1 \ang$ scan step-size.
\Cref{fig:single-slice} shows a single-slice ptychographic reconstruction in comparison to the simulated ground truth electrostatic potential.
Compared with the DPC and parallax reconstructions in~\cref{fig:dpc,fig:parallax}, the improvement ptychography achieves over these methods, by enabling super-resolution set by the largest acquired spatial frequency, is apparent. 
While $k_{\mathrm{max}}$ determines the resolution limit of the reconstruction, the extent of real-space probe-overlap and the dynamic range of the diffraction intensities control the redundancy in the dataset and ultimately the quality of the reconstruction.
Due to the high dose and large $k_{\mathrm{max}}$ used here, ptychography performs better than the other phase retrieval techniques considered here.

\begin{figure*}
    \includegraphics[width=\textwidth]{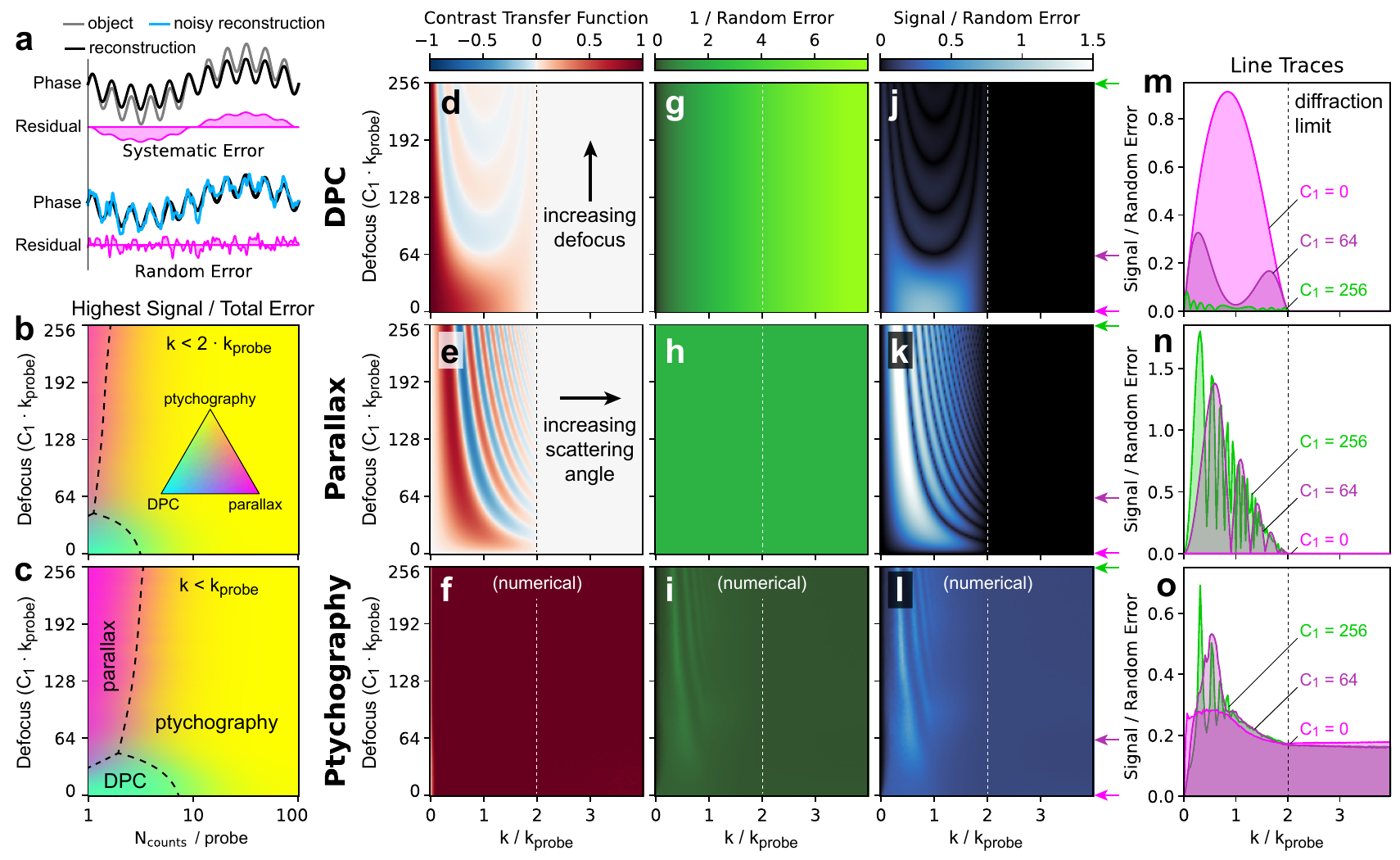}
    \caption{
    Transfer of information of STEM phase-retrieval techniques. 
    a) Schematic representation of systematic and random errors in reconstructions.
    b-c) Ternary ``phase-diagrams'' illustrating the reconstruction technique which achieves the highest signal to total error, as a function of defocus and electron fluence.
    d-f) CTF, g-i) inverse random error, and j-l) signal to random error for DPC, parallax, and GD ptychographic reconstructions, as a function of defocus and spatial frequency.
    m-o) Line traces of the signal to random error plots shown in j-l) for no, moderate, and high defocus values, as denoted by the color-coded arrows.
    }
    \label{fig:CTF}
\end{figure*}

\section{Transfer of Information Comparison} \label{sec:ctf}

To model the accuracy of our STEM phase retrieval methods we estimate values for the contrast transfer function (CTF), the Poisson shot noise error $\Delta \phi^{\mathrm{rand}}$, and systematic reconstruction errors $\Delta \phi^{\mathrm{syst}}$. 
We have implemented a numerical model to measure these values as a function of experimental parameters by reconstructing a pure phase object consisting of white noise (i.e. equal power in all Fourier components).

In our numerical models we use a constant Fourier amplitude of $\phi_0$ = 1, a pixel size of 0.125 \ang, a voltage of 300 kV, a probe convergence semiangle of $k_{\rm{probe}}$ = 1 \ang$^{-1}$, 128 gradient descent iterations, and a field of view comprised of 256$^2$ pixels.
We repeat each numerical test 128 times with both a new random pure phase object and random Poisson noise representing variable electron fluence, averaging the results.
Where possible, these numerical models have been replaced with analytic expressions which were confirmed using the numerical simulations.

Note that white noise phase objects are required to estimate signal transfer beyond the diffraction limit for all phase contrast 4D-STEM methods, since accurately reconstructing the higher spatial frequency terms requires that lower frequency terms be correctly determined, allowing for the relative phase of the overlapping higher frequency terms to be solved. 
Our testing revealed that as long as there was enough probe overlap for the iterative algorithms to converge, none of the tested values depend on the amount of probe sampling -- all measured values asymptotically approach constant values as the STEM probe sampling becomes finer.
Finally, we did not extensively test strong phase objects, instead focusing on the asymptotic signal transfer for weaker phase shifts.

\subsection{Systematic and Random Phase Errors}

The phase contrast CTF is defined as the ratio between the measured output signal (\cref{fig:CTF}a, black) and the simulated phase signal (\cref{fig:CTF}a, gray), typically measured as a function of experimental parameters such as wave aberration coefficients or the probe convergence semiangle $k_\mathrm{probe}$.
CTF values range from -1 to +1, producing negative values when the measured signal is inversely proportional to the input.
Non-unity CTF values represent systematic phase errors inherent to the reconstruction method (\cref{fig:CTF}a, magenta).
Assuming sign-flips due to negative CTF regions are corrected, the residual systematic errors are given by: 
\begin{align}
    \Delta \phi^{\mathrm{syst}}(\boldsymbol{k}) = \left(1 - |\mathrm{CTF}(\boldsymbol{k})|\right) \phi_0 N,
\end{align}
where $N$ is the electron fluence per probe.

The phase error $\Delta \phi^{\mathrm{noise}}$ due to finite electron fluence is estimated by taking the difference of the absolute magnitude of a reconstruction with (\cref{fig:CTF}a, cyan) and without (\cref{fig:CTF}a, black) noise.
The dependence of this phase error on the simulation's signal magnitude $\phi_0$ and total electron dose $N$ can be removed by calculating the normalized random phase error:
\begin{align}
    \Delta \phi^{\mathrm{rand}}(\boldsymbol{k}) = \frac{\sqrt{\mathrm{N}}}{\phi_0} \Delta \phi^{\mathrm{noise}}(\boldsymbol{k}) \label{eq:noise}.
\end{align}

We define the normalized total phase error, equal to the systematic and noise errors combined in quadrature as:
\begin{align}
    \Delta \phi^{\mathrm{total}}(\boldsymbol{k}) &= \frac{1}{\phi_0 N}  \sqrt{ \left(\Delta \phi^{\mathrm{syst}}(\boldsymbol{k})\right)^2 + \left(\Delta \phi^{\mathrm{noise}}(\boldsymbol{k})\right)^2} \nonumber \\
    &= \sqrt{\left(1 - |\mathrm{CTF}(\boldsymbol{k})|\right)^2 + \left(\Delta \phi^{\mathrm{rand}}(\boldsymbol{k})\right)^2 / N} \label{eq:total-error}.
\end{align}

We use this normalized error metric to estimate which phase contrast STEM method produces the lowest phase error, shown in~\cref{fig:CTF}b-c as ternary ``phase diagrams'' for spatial frequencies below the diffraction limit and convergence-angle respectively.
The model predicts that ptychography will produce lower phase errors for higher electron fluences, while DPC and parallax will produce lower errors at lower electron fluences for low and high defocus values, respectively.
This is because at a high electron fluence systematic errors dominate, while at lower fluence the shot noise error dominates.
Dashed lines have been overlaid to delineate the ``phase-boundaries'' between each technique, as a guide to the eye.
For spatial frequencies beyond the diffraction limit, ptychography produces the lowest total error under all conditions as expected.
It should however be noted that at very low electron fluences, the shot noise error in ptychography is high-enough to introduce larger phase errors than the null hypothesis of a zero phase estimate, suggesting a smaller collection angle which is subsequently zero-padded to a desired $k_{\mathrm{max}}$ might be preferable~\citep{sha2023information}. 
Both of these observations suggest that ptychography reconstructions require a minimum electron fluence for high accuracy.
Ptychography also continues to benefit from increasing electron fluence, whereas the error for both DPC and parallax phase measurements asymptotically approaches a constant non-zero value when the accuracy is limited by the systematic errors.

\subsection{Contrast Transfer Functions}

To investigate the information-rich ternary ``phase-diagrams'' shown in~\cref{fig:CTF}b-c, we elaborate on each error component in~\cref{eq:total-error}.
For a DPC measurement, we expect the CTF to be given by the normalized intensity probe autocorrelation function:

\begin{align}
    \mathrm{CTF}_{\mathrm{DPC}}(\boldsymbol{k}) = 
    \mathcal{F}\left[    
        \left|\mathcal{F}^{-1}
            \left[
                A ( {\boldsymbol{k}} )
                \exp \left(
                    - i \pi \lambda |\boldsymbol{k}|^2 C_1
                \right)
            \right] 
        \right|^2
    \right],
\end{align}
where $A(\boldsymbol{k})$ is the probe-forming aperture, and we only consider the defocus $C_1$ contribution to the aberration surface.
This expression is plotted in~\cref{fig:CTF}d as a function of normalized scattering vector $k / k_{\mathrm{probe}}$ and defocus.
\Cref{fig:CTF}d clearly indicates that DPC measurements should be performed as close to zero defocus as possible, and that contrast flipping is possible at large defocus values.

For a parallax measurement, the CTF may be written as the aperture autocorrelation function multiplied by the bright-field TEM CTF~\citep{saxton2000new,meyer2002new,meyer2004new,lupini2010aberration}, equal to:
\begin{align}
    \mathrm{CTF}_{\mathrm{parallax}}(\boldsymbol{k}) = 
    \mathcal{F}\left[    
        \left|\mathcal{F}^{-1}
            \left[
                A ( {\boldsymbol{k}} )
            \right] 
        \right|^2
    \right]
    \sin\left(
        - \pi \lambda |\boldsymbol{k}|^2 C_1
    \right).
\end{align}
This expression is plotted in~\cref{fig:CTF}e. 
The aperture autocorrelation is the ideal signal transfer for a converged probe measurement, and is also known as the ``double-overlap function''~\citep{yang2015efficient}.
The bright-field CTF produces concentric rings referred to as ``Thon rings'', comprising bands of alternating sign with decreasing spacing with increasing defocus~\citep{saxton2000new,meyer2002new,meyer2004new,lupini2010aberration}.
Parallax reconstructions require some defocus to produce any contrast, but note that too much defocus may reduce the measured signal due to coherence limitations.
Recovering the sample's electrostatic potential from parallax measurements additionally requires ``CTF-correction'', where the CTF is estimated so that spatial frequencies with negative CTF can be sign-flipped as in~\cref{subsec:parallax-preprocessing-results}.

Our numerical tests of the CTF for the GD ptychographic reconstruction algorithm shown in~\cref{fig:CTF}f indicate that:
\begin{align}
    \mathrm{CTF}_{\mathrm{ptycho}}(\boldsymbol{k}) = 1,
\end{align}
i.e.~the algorithm correctly recovers the phase and amplitude of all spatial frequencies when reconstructing a white noise object, even those beyond the diffraction limit. 
Note that at the lowest spatial frequencies we estimate CTF values slightly below 1 due to slow convergence speeds. 
Reconstructions of the lowest spatial frequencies in phase contrast TEM have been shown to require a large number of iterations to converge \citep{ophus2012guidelines}.
Our testing suggests that given enough iterations, all spatial frequencies can be fully recovered in a noise-free reconstruction.

\subsection{Random Phase Errors}

The normalized random phase error for DPC and parallax are analytically derived to be given by:
\begin{align}
    \Delta \phi^{\mathrm{rand}}_\mathrm{DPC}(\boldsymbol{k}) &= 
    \frac{ \sqrt{2} k_{\rm{probe}} }{ \pi k } \\
    \Delta \phi^{\mathrm{rand}}_\mathrm{parallax}(\boldsymbol{k}) &= 
    \frac{ \sqrt{2} }{ \pi}.
\end{align}

\Cref{fig:CTF}g-h plots the inverse of these expressions, respectively, while~\cref{fig:CTF}i shows the numerical estimate for ptychography. 
We plot the inverse errors so that the regions with a higher signal-to-noise ratio (SNR) can be easily identified. 
Overall, ptychography shows increased phase error relative to both DPC and parallax.
This is especially prevalent at the lowest spatial frequencies, indicating that ptychography may produce more noise in the lowest spatial frequencies than DPC. Note this is not the case with parallax imaging as its CTF approaches zero at the lowest spatial frequencies. 
Interestingly, some spatial frequencies show reduced random error for ptychographic reconstructions using a defocused probe, with the pattern directly following the CTF calculated for parallax. 
However, these ``Thon rings'' of increased SNR are considerably sharper than those of parallax or BF TEM, roughly equal to the parallax CTF to the fourth power.

\Cref{fig:CTF}j-l show the estimated ratio of signal to random errors as $|\mathrm{CTF}| / \Delta \phi^{\mathrm{rand}}$ for DPC, parallax, and GD ptychography, with line traces shown in~\cref{fig:CTF}m-o respectively.
The DPC SNR shown in~\cref{fig:CTF}j and m approximately recovers the profile estimated using a different method~\citep{yang2015efficient}.
DPC produces the highest signal-to-noise close to $k = k_\mathrm{probe}$ and at zero defocus, falling off at higher and lower spatial frequencies.

The parallax SNR shown in~\cref{fig:CTF}k and n approaches the ideal converged probe signal transfer as the defocus is increased, though with roughly half of the spatial frequencies transferred with rapid oscillations.
As mentioned above, care must also be taken in choosing a defocus high enough to produce significant signal transfer without introducing coherence limitations. 

The GD ptychography SNR shown in~\cref{fig:CTF}l and o is much flatter than that of either DPC or parallax. 
Ptychography is also the only method which can recover signal beyond the diffraction limit, $k > 2 k_\mathrm{probe}$.
Additionally, ptychography does not produce zero SNR at any spatial frequency, showing that it models the signal transfer in a much more accurate manner.
This model also indicates that using a defocused probe for ptychographic imaging can somewhat improve the signal transfer relative to imaging in-focus, at least for low electron fluences, though this is a relatively minor difference.
However, note that the overall ptychography SNR is approximately half of the mean value for DPC and parallax below the diffraction limit.

\subsection{CTF and Phase Error Models Limitations}

Finally, we discuss limitations of our CTF and phase error models. 
First, we warn readers against over-interpreting any specific numbers given by these models, but rather try to understand the general trends.
In particular, real experimental samples are never equivalent to white noise pure phase objects.
Atomic potentials fall off as $k^{-2}$ for high spatial frequencies, and thus we expect reaching super-resolution in GD ptychography will require much larger electron fluences than those given above.
There may also be strong correlations in the electrostatic potential signals, particularly for crystalline Bragg samples, which may influence the resulting measurements.
We have also not explicitly considered direct, non-iterative ptychographic methods such as single sideband ptychography, though we expect it to fall roughly inline with the DPC analysis considered above.

Each of the methods we have considered have other potential sources of error.
For example DPC relies on accurate rotation calibration (\cref{subsec:dpc-preprocessing-results}).
Parallax requires accurate cross-correlation estimates between all of the virtual images (\cref{subsec:parallax-preprocessing-results}), which can be challenging at low electron fluence, though this can partly compensated using fiducial markers.
Ptychography requires accurate calibration of all of the experimental parameters, and ptychographic reconstructions may not converge if these calibrations are not sufficiently accurate (\cref{subsec:ptycho-preprocessing}).
Ptychographic reconstructions can also fail if the data is not recorded with sufficient redundancy.

We have also ignored the influence of amplitude variations in the object due to electrons scattered outside the range of the detector, or the effect of very strong scattering.
Thicker samples too will violate the projected potential assumption which is implicit in DPC, parallax, and single-slice ptychographic reconstructions.
Despite these limitations, we hope the above analysis will give some insight into the transfer of information of STEM phase retrieval techniques.

\section{Experimental Comparison} \label{sec:experimental-results}

The previous sections use simulated datasets to illustrate the differences between STEM phase retrieval algorithms.
As emphasized in the introduction, our phase retrieval algorithm implementations and specifically the regularization constraints introduced in~\cref{sec:experimental-considerations} are designed to be robust against common experimental artifacts. 
To this end, in this section we test our implementations on both materials science and biological samples. 

\subsection{Gold Nanoparticles} \label{subsec:exp-gold}

\begin{figure*}
\includegraphics[width=\textwidth]{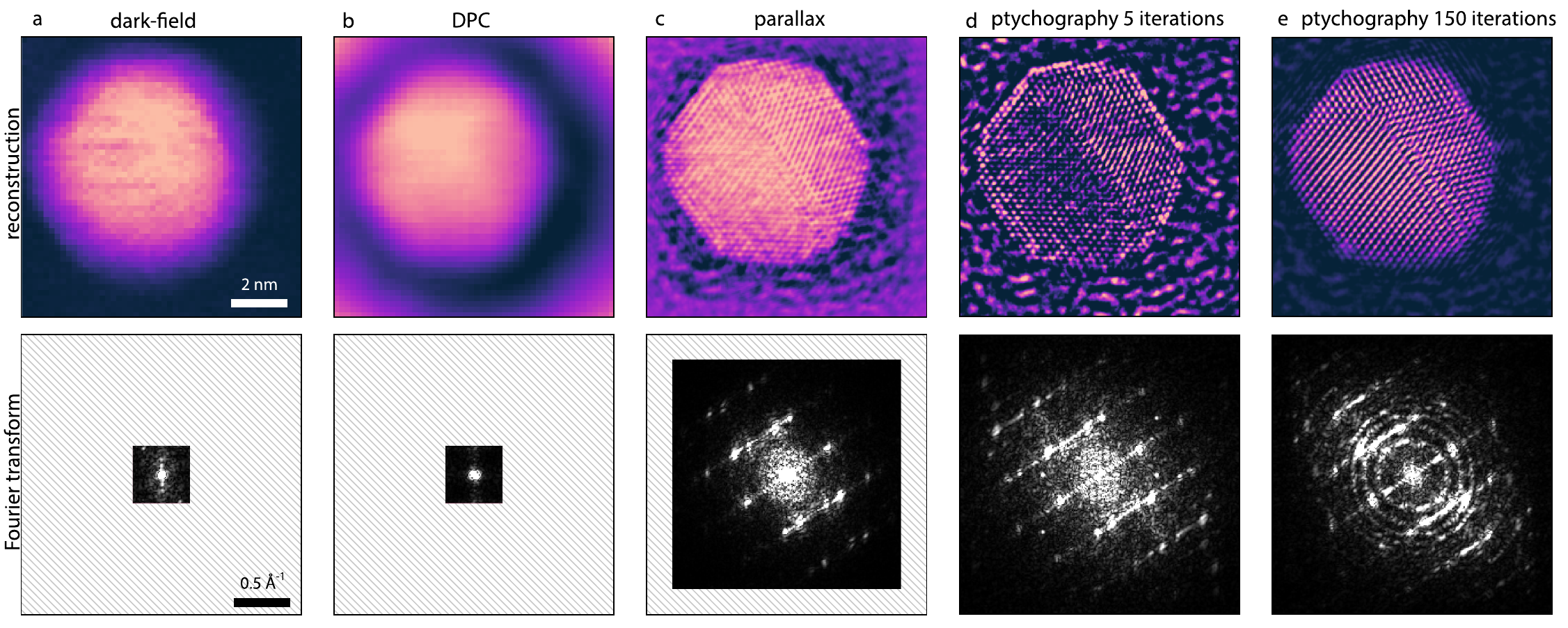}
\caption{Experimental comparison of phase retrieval STEM using gold nanoparticles.
a) Annular dark-field imaging and b) DPC reconstructions are limited by real-space scan sampling.
c) Parallax reconstruction can correctly account for probe aberrations and reveal atomic features beyond the scan sampling frequency.
d) Ptychographic reconstruction after 5 iterations captures high spatial frequencies but misses low frequency signal and accentuates noise. 
e) After 150 iterations, the reconstruction converges, capturing low and high spatial frequencies providing the highest resolution reconstruction.}
\label{fig:gold}
\end{figure*}

We highlight the strengths and limitations of STEM phase retrieval techniques for materials characterizations using a defocused experimental dataset of gold nanoparticles on amorphous carbon film.
\Cref{fig:gold} illustrates virtual dark-field imaging as-well as different reconstructions using DPC, parallax, and ptychography. 
These images are different reconstructions from the same experimental 4D-STEM dataset. 
Annular dark-field and DPC (\cref{fig:gold}a-b) are limited both by aberrations in the incident beam and real-space scan sampling. 
Thus for a defocused dataset with a 2\ang~sampling, these methods provide low-resolution images, where it is not possible to resolve any atomic features. 
Dark-field imaging, which has monotonic contrast with the projected sample potential, does accurately capture the outline of the gold nanoparticle. 
By contrast DPC, which has poor signal-to-noise for low-spatial frequencies (\cref{sec:ctf}), is not able to accurately describe the contrast difference between the gold nanoparticle and the amorphous carbon support. 

\Cref{fig:gold}c highlights the advantages of parallax imaging for defocused datasets such as this one.
Unlike dark-field and DPC, parallax reconstructions can correctly account for the aberrations of the incoming probe, computationally bringing the dataset in-focus (\cref{sec:parallax}).
In combination with subpixel accurate cross-correlation shifts and kernel-density estimation upsampling, this enables the recovery of high-resolution features, such as gold Bragg peaks. 

For this dataset, the highest resolution reconstruction is enabled using GD ptychography. 
\Cref{fig:gold}d-e show the reconstructed projected potential after 5 and 150 iterations, respectively. 
As discussed above, low spatial frequencies are particularly slow to converge in ptychography, so the largest difference between~\cref{fig:gold}d-e is that the thickness contrast from the gold nanoparticle has not converged yet. 
The improved convergence is further apparent by the appearance of ``Thon rings'' in the Fourier transform of the image.
The remaining improvements are enabled by the ability to enforce various object, probe, and position constraints~(\cref{subsec:object-regularization,subsec:probe-regularization,subsec:positions-regularization}).
First, we have applied position correction to remedy the scan distortion visible in the dark-field image.
Second, the probe was constrained at every iteration by replacing its amplitude with a vacuum probe aperture measurement while smoothing its phase to a low-order aberration surface expansion. 
Finally, we also regularized the object by forcing it to be a positive potential object and applying a small amount of low-pass Fourier filtering to reduce the high-frequency noise.
\Cref{fig:gold} highlights the advantages of ptychography for high-resolution characterization. 
Importantly, our reconstruction model is flexible, allowing one to introduce various object, probe, and position regularization to arrive at a physical solution.  

\subsection{2D Hexagonal Boron Nitride} \label{subsec:exp-hbn}

\begin{figure}
\includegraphics[width=\linewidth]{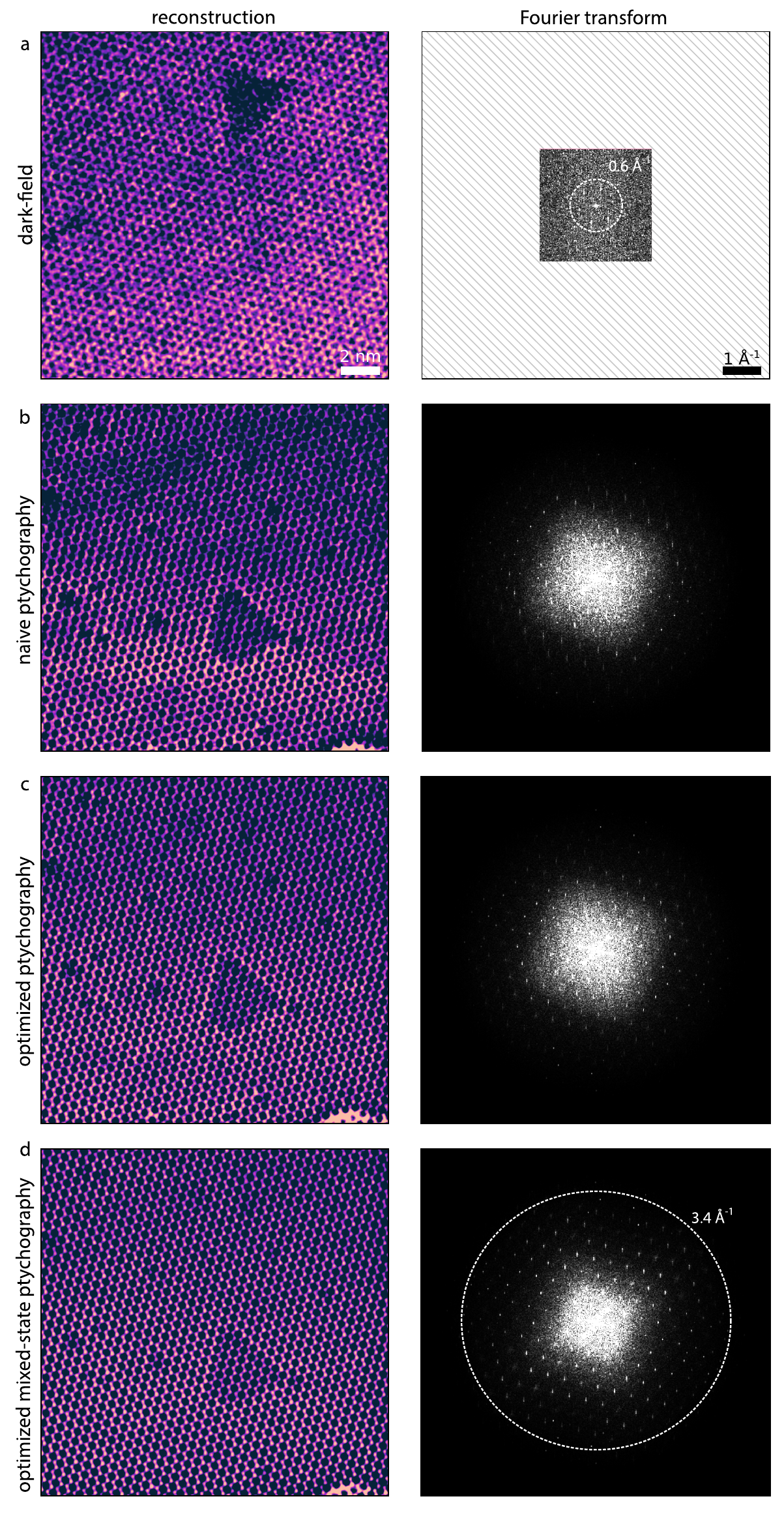}
\caption{Experimental comparison of ptychographic reconstructions of hBN.
a) Dark-field imaging is limited by the probe size at 80kV with a 23 mrad convergence angle. 
b) Ptychography provides much higher signal-to-noise. 
c) Optimizing the calibrations and regularizing the reconstruction leads to an improved phase image with sharper atomic features and fewer low spatial frequency artifacts.
d) The reconstruction is further improved with mixed state ptychography.}
\label{fig:hbn}
\end{figure}

The advantages of ptychography for materials-science samples are further explored by characterizing a few-layer hexagonal boron nitride (hBN) sample containing vacancy defects~\citep{meyer2009selective,tran2016robust}.
\Cref{fig:hbn}a shows a dark-field STEM image of the few-layer hBN sample. 
While it is possible to see lattice contrast and the large triangular vacancy defect cluster in the sample, it is difficult to resolve individual atoms, including any single point defects in the sample.
The Fourier transform illustrates how limited in resolution the dark-field image is at this convergence angle, voltage, electron dose, and sampling. 

\Cref{fig:hbn}b-d compare the dark-field image to ptychographic reconstructions of a similar field of view of the same sample with the same microscope alignment. 
In addition to the signal-to-noise benefits of ptychography, the maximum scattering angle determines the resolution. 
Here, we pad the data in diffraction-space to reconstruct high-frequency information that would otherwise be aliased in the reconstruction. 
We highlight the differences between~\cref{fig:hbn}b, a reconstruction without any regularization applied during the reconstruction, and~\cref{fig:hbn}c, an optimized and regularized reconstruction(see ~\cref{subsec:object-regularization}).
Comparing these two images, the systematic low-frequency error is reduced in the optimized reconstruction, as we would expect from prior knowledge of the sample that the contrast across the field of view be nearly uniform.
Moreover, the atomic columns are sharper, which is evidenced in both the real-space image and the absence of ``streaking'' in the Fourier transform due to scan noise. 
Both the low and high spatial frequencies can be further improved in the reconstruction using a mixed-state model with 2 probes (\cref{fig:hbn}d), which will be further discussed in~\cref{sec:mixed-state_ptycho}.
\Cref{fig:hbn}d shows that the ptychographic reconstruction reaches an information limit of 29~pm (3.4 \ang$^{-1}$), which is much higher than the resolution achieved with dark-field imaging with these parameters, making this a ``super-resolution" reconstruction.

\subsection{Vitrified Virus-like Particles} \label{subsec:exp-vlp}

Traditionally, imaging of biological samples has been performed with parallel illumination TEM or with stained specimens to help guard against beam-induced damage. 
However, recent work has highlighted the possibilities for STEM phase retrieval to characterize vitrified hydrated proteins, organelles and whole cells~\citep{kucukoglu2024low, seifer2024optimizing, yu2022dose,yu2024dose}.
In particular, STEM phase retrieval methods produce high contrast micrographs, which have been proposed to be especially beneficial for very small proteins or thick biological assemblies~\citep{rez2016exploring, seifer2024optimizing}.
In this section, we highlight how the dose-efficiency of the phase retrieval methods we have outlined above, which have predominately been used in the physical sciences, can be applied to study biological systems. 

\begin{figure}
\includegraphics[width=\linewidth]{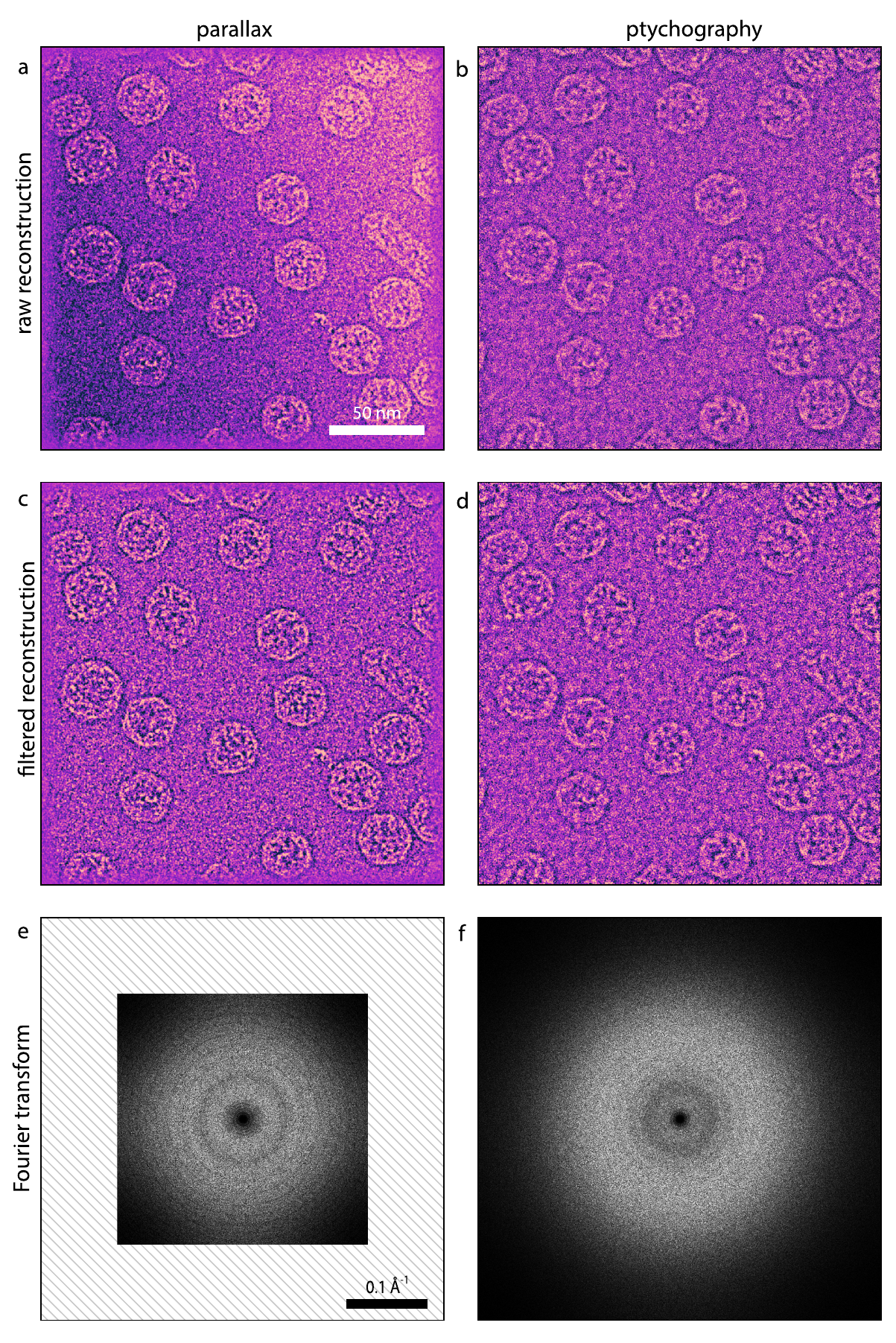}
\caption{Phase retrieval 4D-STEM for biological samples.
a) Parallax and b) ptychographic reconstructions of VLPs. 
c) High and d) low frequency filtering improves contrast in these images.
e-f) Corresponding Fourier transforms of c and d respectively.
}
\label{fig:vlp}
\end{figure}

\Cref{fig:vlp} plots the parallax and ptychographic reconstructions of a sample composed of virus-like particles (VLP) embedded in vitreous ice. 
For low electron dose experiments such as these, it can be challenging to properly converge the parallax reconstruction which relies on the accurate alignment of virtual bright-field images. 
This can, however, be achieved by running initial iterations at high real-space binning values, which are then used to seed lower bin alignments iteratively. 
Due to the sub-pixel accuracy of the calculated cross-correlation shifts, we use KDE upsampling (\cref{subsec:parallax-preprocessing-results}) to obtain a high-resolution phase reconstruction of the VLPs. 
While the reconstruction in \Cref{fig:vlp}a shows some low-spatial frequency artifacts, these can be removed with filtering (\Cref{fig:vlp}c). 

For the ptychographic reconstruction, we optimize parameters to produce the ptychographic potential image in \Cref{fig:vlp}b. 
For physical-sciences samples we often constrain the object to be a positive potential object.
However, for biological samples plunge-frozen in vitreous ice, we do not use the positivity constraint due to the outline of small negative phase shifts due to the gap between the protein and water molecules~\citep{shang2012hydration}. 
This may lead to a noisier reconstruction, but allows us to recover the structure of weakly scattering objects that may fall into the noise floor and otherwise be cut off with a strong positivity constraint. 
Similar probe constraints are used for biological and materials samples.
After reconstruction, the ptychographic reconstruction can also be filtered to improve contrast (\Cref{fig:vlp}c), which provides an image with similar resolution to parallax imaging as highlighted by the Fourier transforms of the filtered reconstructions in \Cref{fig:vlp}e-f.

It can be challenging to compare the resolution of the parallax and ptychography in~\cref{fig:vlp} from 2D micrographs alone. 
While the ptychographic potential image has higher sampling than the parallax reconstruction, the cone-weighted Fourier transforms highlight that there is little transfer of information beyond the semi-convergence angle in parallax.
Further work is required to compare the trade-offs between these two techniques for characterizing biological structures~\citep{kucukoglu2024low}. 

\section{Methods} \label{sec:methods}

STEM simulations were preformed using the \texttt{abTEM} package \citep{madsen2021abtem}. 
The double-walled CNT structure was constructed using the \texttt{atomic simulation environment} (ASE) package~\citep{larsen2017atomic}.
The iterative DPC simulations shown in~\cref{fig:dpc} were performed at 80kV with a 25 mrad convergence semi-angle, at zero defocus, using 12 frozen-phonon configurations with a displacement standard deviation of 0.075 \AA. 
The scan step size and reciprocal sampling were 0.25 \AA~and 2.5 mrad respectively. 
A rotation of 17$^\circ$ and transpose, as well as finite dose of $10^5$e$^-$/\AA$^2$ were added to the dataset. 
The parallax simulations shown in~\cref{fig:parallax} used the same parameters but 0.5\AA~step size, 0.7 mrad reciprocal-space sampling and 20 nm of defocus. 
The single-slice ptychography simulations shown in~\cref{fig:single-slice,fig:projection-sets} used the same parameters but a 0.5\AA~step size, 2.5 mrad reciprocal-space sampling, 15 nm of defocus, and finite dose $10^6$e$^-$/\AA$^2$.
The scan step sizes and finite electron doses in~\cref{fig:dose-overlap} varied between 0.5 \AA~ to 3.0\AA~and $10^3$e$^-$/\AA$^2$ to $10^7$e$^-$/\AA$^2$ respectively.
Finally, the tilt series shown in~\cref{fig:overlap-tomo} was comprised of (30) 20 tilt angles from ($\pm$90$^\circ$) $\pm$60$^\circ$ in 6$^\circ$ increments for the case (without) with a missing wedge, and finite dose of $10^4$e$^-$/\AA$^2$.

The SrTiO$_3$ structure was constructed using ASE from a unit-cell file taken from the \texttt{materials project}~\citep{jain2013commentary}.
The mixed-state and multi-slice ptychography simulations shown in~\cref{fig:mixed-state,fig:multi-slice} were performed at 200kV with a 20 mrad probe, 10 nm of defocus, 12 frozen-phonon configurations with a displacement standard deviation of 0.1 \AA, and sampling of 0.5 \AA~and 1.8 mrad in real and reciprocal-space respectively.
The incoherent mixed-state dataset was obtained by adding the intensities of two independent simulations using the simulated probes shown in~\cref{fig:mixed-state}b with a 3:1 ratio and a finite electron dose of $10^5$e$^-$/ \AA$^2$.
The thickness of the SrTiO$_3$ sample used in multi-slice simulations was varied by repeating the structure along the beam direction matching the thicknesses described in the text, and using a finite electron dose of $10^4$e$^-$/ \AA$^2$.

The gold experimental dataset in~\cref{fig:gold} was taken on the TEAM I microscope at the National Center for Electron Microscopy, a modified FEI Titan double-aberration-corrected microscope operating at 300kV with a probe defined by a 20.3 mrad convergence angle.
A step size of 2 \AA~ was used, and the probe was defocused to a size of about 1.5 nm.
The diffraction data was energy filtered with a Continuum Gatan imaging filter (GIF) and then acquired on a Gatan K3 camera using an electron dose of  8$\times 10^4$e$^-$/ \AA$^2$. 
The data was hot-pixel filtered before reconstruction. 
The data was binned by 8 before reconstruction such that the reciprocal pixel size was 0.5 mrad with a maximum collection angle of 27 mrad.
The parallax reconstruction was reconstructed down to bin 1 and was then upsampled with a KDE factor of 4. 
The ptychographic reconstruction used a gradient descent approach for 150 iterations with a batch size of 156 measurements and a step size of 0.1. 
The probe was constrained by fixing the probe amplitude with a vacuum measurement taken during the same session, and the phase was smoothed by fitting a low-order aberration surface expansion. 
The positions were refined during each iteration using a gradient step on the intensity measurements to create a displacement vector for each probe position with a step size of 0.005.
The object was constrained to be a positive potential object and was low pass filtered with a Butterworth filter at a critical frequency of 1.5\AA$^{-1}$ and an order of 2.

The hBN experimental dataset in~\cref{fig:hbn} was imaged on the TEAM I microscope, operated at 80kV with a probe convergence angle of 22.9 mrad. 
The few layer hBN sample was fabricated via tape exfoliation from bulk hBN and dry transfer to a holey SiNx grid. The sample was then baked at 400$^\circ$C for 15 minutes immediately before loading into the TEM. 
A step size of 0.69 \AA~ was used, and the probe was defocused to a real space size of about 8 \AA.
The diffraction data was acquired on the Dectris Arina camera, with an electron dose of approximately 2$\times 10^5$e$^-$/ \AA$^2$. 
The data was flat-field corrected and then hot-pixel filtered before reconstruction. 
The dark-field image was acquired with a similar dose to the scanning diffraction dataset. 
The naive ptychographic reconstruction was initialized with a vacuum probe acquired from the same session. 
The data, which had a reciprocal pixel size of 0.5 mrad and a maximum collection angle of 48 mrad, was padded by a factor of 4. 
The gradient descent algorithm was used with 400 steps of size 0.3.
The optimized ptychographic reconstruction was run with the same initial probe, padding, number of iterations, and step size.
However, the real space and reciprocal calibration was optimized with the Bayesian Optimizer.
During the reconstruction, the Fourier amplitude of the probe was constrained, and position refinement was turned on.
The object was also reconstructed using a mixed-state model with 2 probes using otherwise identical constraints.
Only the first mixed-state probe is constrained.

The VLPs shown in~\cref{fig:vlp} were coat proteins of bacteriophage PP7 self-assembled during recombinant expression in E. coli, plunge-frozen on TEM grids using R1.2/1.3 mesh 300 UltrAuFoil grids.
The data was acquired on an uncorrected Thermofisher Scientific Krios G4 with a dedicated cryo-stage operated at 300kV using an EMPAD detector.
The step size was 1.3 nm, the semi-convergence angle was 4 mrad, and the probe was defocused to about a 15 nm diameter.
The data was taken so that the central beam filled most of the detector – the maximum collection angle was about 4.8 mrad with a pixel size of 0.08 mrad.
The dose was approximately 41 e$^-$/ \AA$^2$, but roughly 21 e$^-$/ \AA$^2$ extra dose was lost to detector dead time.
The camera frame-rate was set to 1kHz and the detector dead time is 0.5 ms per acquisition.
To keep the reconstruction from diverging, the binning in parallax was never reduced below 16, but 8 iterations were run at the minimum bin to help converge the reconstruction. 
A KDE upsample factor of 4 was used.  
After the reconstruction, the object was high pass filtered with a Butterworth filter at a critical frequency of 0.005\AA$^{-1}$ and an order of 2 and low pass filtered with a Butterworth filter at critical frequency of 0.15\AA$^{-1}$.
Ptychographic reconstructions were performed with a gradient descent approach with a batch size of 10 and a step size 0.1. 
10 iterations were run. 
For this reconstruction, the probe was also constrained by fixing the probe amplitude with a vacuum measurement taken during the same session, and the phase was smoothed by fitting a low-order aberration surface expansion. 
The positions were refined during each iteration using a gradient step on the intensity measurements to create a displacement vector for each probe position with a step size of 0.005.
After the reconstruction, the object was Butterworth filtered with the same critical frequencies as the parallax reconstruction. 
    
\section{Summary} \label{sec:summary}

We have outlined the theory and implementation of various phase retrieval STEM techniques, as well as compared the techniques numerically and experimentally.
The optimal reconstruction technique will depend on the sample, experimental parameters, and specific scientific question. 
Iterative DPC works best for in-focus datasets, with the resolution limited by the Nyquist frequency of the scan-step size. 
Nonetheless, this is the least computationally expensive and most straightforward technique, and iterative DPC may provide sufficient information for many samples. 
Parallax reconstructions require only slightly increased computation over iterative DPC, but enable resolution higher than the scan sampling Nyquist limit, and can provide accurate estimates of low-order aberration coefficients.
Finally, ptychographic techniques provide the highest resolution reconstructions, at increased computational cost and reconstruction parameter complexity.

We further provide extensive discussion on algorithmic and experimental considerations, including STEM-specific regularizations and data acquisition recommendations, which we hope will be useful to the community.
Common extensions of electron ptychography, including mixed-state, multi-slice, and joint ptychographic tomography formalisms are presented in the appendices using a consistent framework, enabling practitioners to directly compare and extend the implementations.

\begin{appendices}

\section{Appendices}

\begin{figure*}
\includegraphics[width=\textwidth]{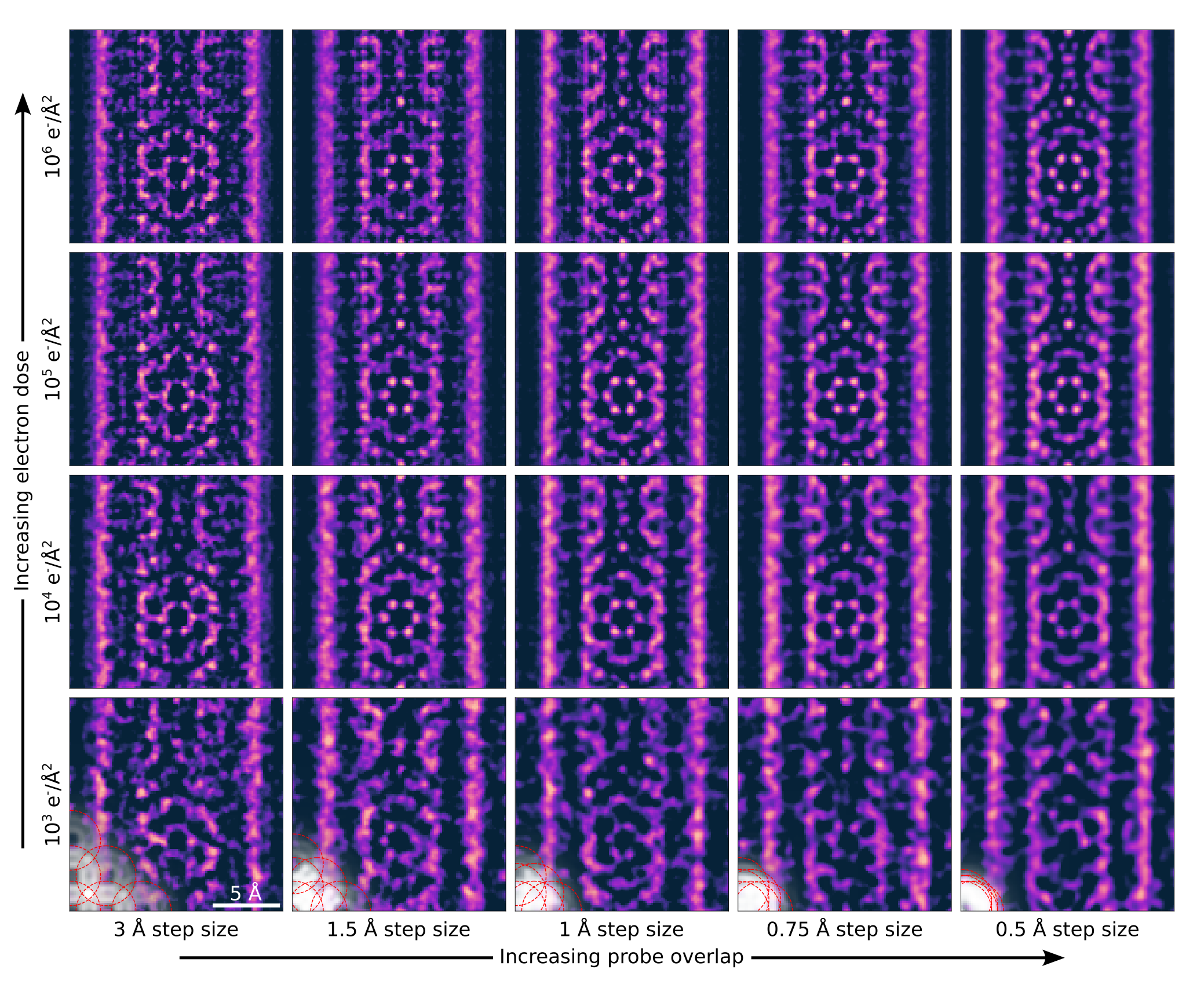}
\caption{
Dependence of single-slice ptychographic reconstruction quality on finite electron dose (rows) and probe overlap (columns).
Decreasing probe overlap results in high-frequency artifacts, while decreasing electron dose results in loss of atomic features.
}
\label{fig:dose-overlap}
\end{figure*}

\section{Algorithmic Considerations} \label{sec:algorithmic-considerations}

\subsection{Probe Overlap and Finite Dose}  \label{subsec:probe-overlap-dose}
  
There are two ways to increase probe-overlap, while maintaining the same real- and reciprocal-space sampling:
i) decreasing the scan step-size between adjacent probes, or
ii) increasing the size of the real-space probe, e.g. by defocusing further or using a smaller probe convergence angle.
The former strategy is the most straight-forward, but necessitates larger acquisition times and dataset sizes.
The latter strategy can be quite effective however is limited both in using larger defocus, which requires larger region-of-interest probe arrays to avoid wrap-around artifacts, and smaller convergence angles, which change the effective transfer of information.

\Cref{fig:dose-overlap} illustrates the dependence on reconstruction quality by starting with the same dataset as~\cref{fig:single-slice} (shown on the top-right tile) and decreasing the electron dose moving vertically down and decreasing the probe overlap moving horizontally to the left.
While the reconstruction quality deteriorates along both these axes, it is interesting to note the difference in the two failure modes.
Decreasing the probe overlap at high dose (top row moving left) results in high-frequency artifacts, while decreasing the electron dose at large overlap (rightmost column moving down) results in substantial loss of atomic features.
It should also be noted that achieving both high electron-dose and large probe overlap is often experimentally challenging for dose-sensitive samples.

\begin{figure*}
\includegraphics[width=\textwidth]{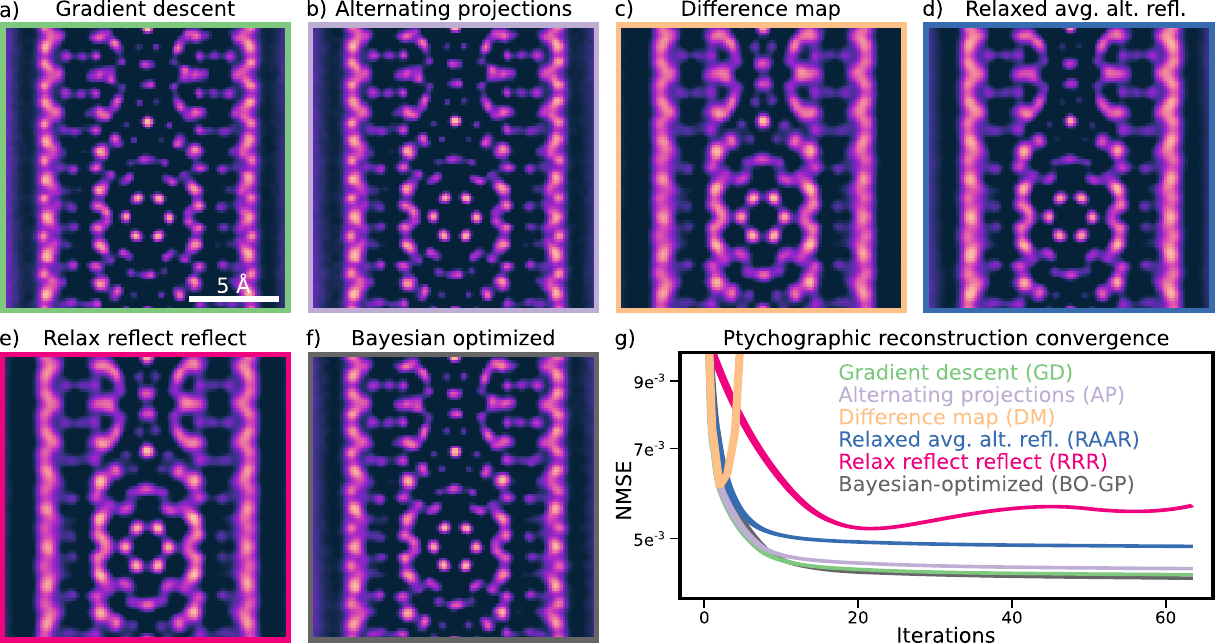}
\caption{
a-f) Comparison between various ptychographic reconstruction algorithms.
For diverging algorithms such as c) DM, the lowest-error reconstruction is displayed instead.
g) Convergence comparison, illustrating how Bayesian optimization can be used to tune the proximal gradient hyperparameters.
}
\label{fig:projection-sets}
\end{figure*}

\subsection{Reconstruction Algorithms Comparison}

Electron ptychography is a high-dimensional, non-convex inverse problem over the complex numbers.
As such, rigorous proofs of convergence and comparisons between different algorithms are few and far between.
In this section, we instead comment on the use of batching in the stochastic gradient descent solver, and various well-known projection-set algorithms based on empirical observations.

First, note that when the batch size in~\cref{eq:gd-psi,eq:gd-update-o,eq:gd-update-p} is equal to a single probe position,  $\mathcal{M}=1$, the stochastic gradient descent algorithm reduces to a regularized version of the well-known extended ptychographic iterative engine (e-PIE)~\citep{maiden2009improved,maiden2017further}, and when $\mathcal{M}=M$ the algorithm is no longer stochastic.
For simulated datasets with high electron dose, we find that using a vanishing normalization parameter, $\alpha \to 0$, achieves the lowest error.
For low-dose experimental datasets, we almost always observe better behavior as $\alpha \to 1$.
While in general we find that using as many probe positions as the available memory permits, $\mathcal{M} \to M$, appears to be both more accurate as well as computationally more performant, we note that using fewer probe positions per iteration allows the algorithm to converge the probe faster.
This is particularly important for mixed-state ptychography, which we elaborate further in~\cref{sec:mixed-state_ptycho}. 

The scalar parameters $\left(a,b,c\right)$ introduced in~\cref{eq:proj-psi,eq:gd-update-o,eq:gd-update-p} describe a whole family of proximal gradient algorithms, with specific values encoding well-known named algorithms such as alternating projections (AP, $a=0,b=1,c=1$)~\citep{neumann1950functional}, difference-map  (DM, $a=-1,b=1,c=2$)~\citep{elser2003phase}, relaxed averaged alternating reflections (RAAR, $a=1-2\gamma,b=\gamma,c=2$)~\citep{luke2004relaxed}, and relax reflect reflect (RRR, $a=-\gamma,b=\gamma,c=2$)\citep{elser2017complexity}.
We make the following observations, summarized in~\cref{fig:projection-sets}:

\begin{enumerate}[i.) ]
    \item The AP (\cref{fig:projection-sets}b) or ``error-reduction'' algorithm is very similar to the gradient descent algorithm when $\mathcal{M}=M$ and $\beta=1$ (\cref{fig:projection-sets}a).
    \item While the DM algorithm (\cref{fig:projection-sets}c), tends to diverge for most datasets we have investigated, variants of the algorithm with relaxation parameters namely RAAR (\cref{fig:projection-sets}d) and RRR (\cref{fig:projection-sets}e) show better convergence behaviour. 
    \item The space of parameters defined by $\left(a,b,c\right)$ lends itself well to dataset-specific hyper-parameter tuning.
    \Cref{fig:projection-sets}g uses Bayesian optimization with gaussian processes to arrive at a set of parameters with better convergence properties $\left( a=-0.34,b=0.47, c=1.1\right)$.
    This optimization is further discussed in \cref{subsec:parameter-tuning}.
\end{enumerate}

Finally, we note that in all the experimental datasets we have encountered so far - we find stochastic GD reconstructions to be superior to proximal gradientmethods, mostly due to the flexibility afforded by the choice of batch-size.

\subsection{Raster-scan Pathology} \label{subsec:grdding}
The 2D raster-scan pattern diffraction intensities are typically acquired on during 4D-STEM experiments results in a periodic ``gridding'' artifact, also called raster scan pathology~\citep{fannjiang2019raster,fannjiang2020blind}.

First, note that this arises only in the case of ``blind'' ptychography -- i.e. when both the probe and the object need to be solved simultaneously.
It follows that one of the ways to avoid the gridding artifacts is using perfect knowledge of the illuminating probe and only solve for the reconstructed object.
While this is not practical for experimental datasets, it can be approximated by recording a high-quality vacuum probe measurement to use as the aperture, and estimating the aberration coefficients using parallax as discussed in~\cref{sec:parallax}.
While this can indeed help converge the reconstruction, in particular for low-dose datasets, we note that any misalignment of the vacuum probe measurement and the experimental datasets will result in model-mismatch which can also result in gridding.

Gridding can be suppressed with increased data redundancy.
As discussed in~\cref{subsec:probe-overlap-dose}, this can be achieved using either fine scan-step sampling or larger probe sizes, e.g. by introducing defocus.
As discussed above, we warn the reader about the latter strategy as the windowed object patch size, set by the detector pixels and maximum scattering angle, needs to be large enough to comfortably fit the long probe tails.
Periodic boundary conditions imposed by the windowed Fourier transforms result in self-interaction of the probe which again introduce model-mismatch and ultimately gridding artifacts.

Gridding can also be partly ameliorated by using small batch-sizes, which allow the probe to update more frequently.
This gives the algorithm an opportunity to self-correct model-mismatch and climb out of local minima.
Taking this a step further, mixed-state ptychography (\cref{sec:mixed-state_ptycho}) can be used to include a sacrificial low-intensity probe to absorb these artifacts.

\begin{figure*}
\includegraphics[width=\textwidth]{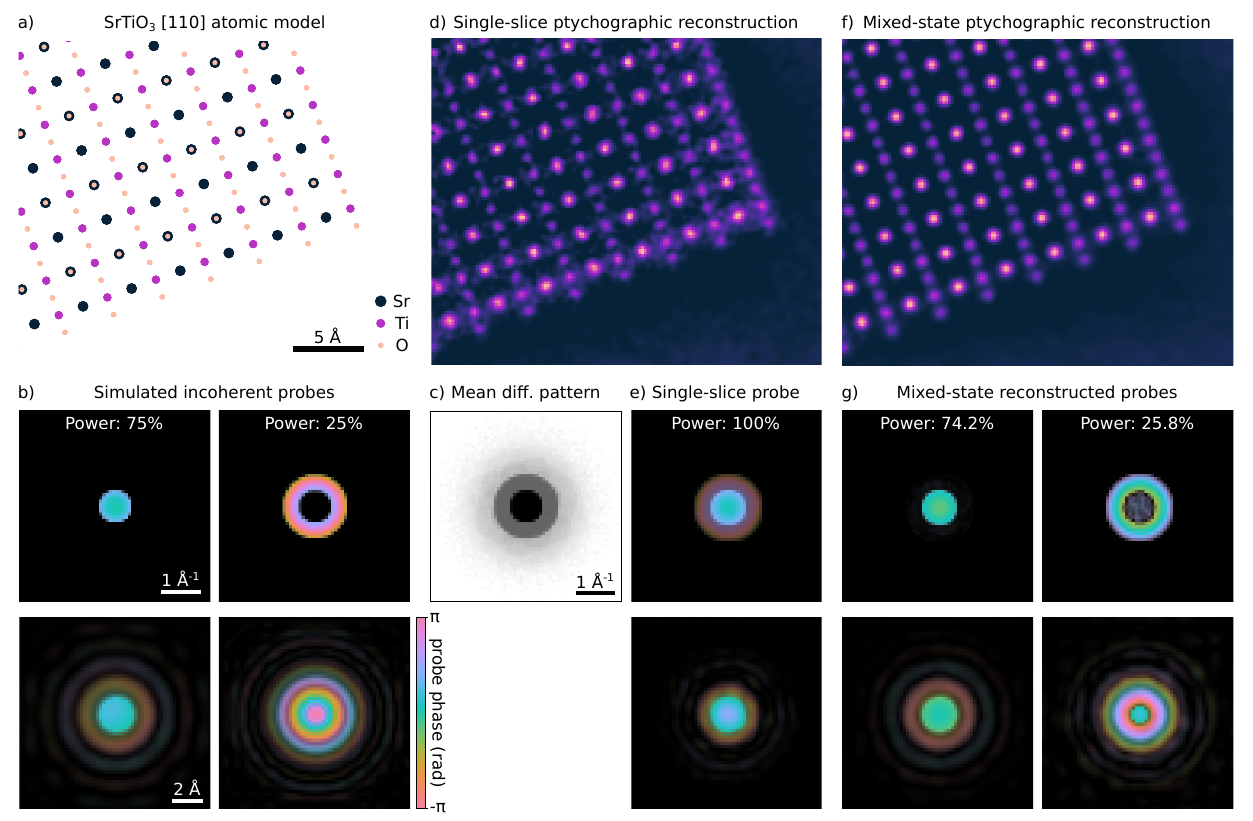}
\caption{
Mixed-state ptychographic reconstruction of simulated STO [110] slab a-c) showing
d-e) the breakdown of single-slice ptychography on datasets with partial coherence, and
f-g) the much better fidelity reconstruction using mixed-state ptychography with two orthogonal probes.
}
\label{fig:mixed-state}
\end{figure*}

\section{Mixed-state ptychography} \label{sec:mixed-state_ptycho}

The formalism developed in~\cref{sec:single-slice_ptycho} is capable of reconstructing the probe aperture and aberrations for coherent illumination.
However, the transfer of information in 4D-STEM, and thus the quality of ptychographic reconstructions, is often limited by imperfections in the imaging system resulting in partial temporal and spatial coherence~\citep{dwyer2010measurement}.
\Cref{fig:mixed-state} illustrates this breakdown for a synthetic dataset of a STO [110] slab (\cref{fig:mixed-state}a), simulated by incoherently adding diffraction intensities simulated with a circular aperture between 0-10 mrad (\cref{fig:mixed-state}b, left) and an annular aperture between 10-20 mrad (\cref{fig:mixed-state}b, right) with $75\%$ and $25\%$ weights respectively.
The single-slice ptychographic reconstruction shown in~\cref{fig:mixed-state}d-e, attempts to inject the partial coherence information in the object while using a single probe aperture, which degrades the reconstruction quality. 

While one approach to remedy this could be to adjust our forward operator to include parametric models of partial coherence, it has been shown that modeling the probe as a linear combination of pure quantum states can effectively capture partial coherence and provides greater flexibility~\citep{thibault2013reconstructing,chen2020mixed}.
In order to capture this, we modify our forward operator in~\cref{eq:overlap-projection,eq:model-intensity} to read:

\begin{align}
    \psi_m^{(l)}(\boldsymbol{r}) &= \Pi_O\left[\mathcal{P}^{(l)},\mathcal{O}\right] = \mathcal{P}^{(l)}(\boldsymbol{r}) \mathcal{O}(\boldsymbol{r}-\boldsymbol{R}_m) \label{eq:overlap-projection-mixed} \\
    J_m(\boldsymbol{k}) &= \sum_l \left| \mathcal{F}\left[\psi_m^{(l)}(\boldsymbol{r})\right] \right|^2 \label{eq:model-intensity-mixed},
\end{align}
with the Fourier projection and adjoint operators in~\cref{eq:fourier-projection,eq:gd-psi,eq:gd-update-o,eq:gd-update-p,eq:proj-psi,eq:proj-update-o,eq:proj-update-p} applying \textit{mutatis mutandis}.

\Cref{eq:model-intensity-mixed} holds for $L$ orthogonal probe functions $\left\{ \mathcal{P}^{(l)}\right\}$, a constraint which is not \textit{a-priori} guaranteed by the adjoint operator and must instead be imposed at every iteration.
We implement a computationally-efficient variant of the orthogonal probe relaxation (OPR) algorithm~\citep{Odstrcil2016}, which uses the eigenvectors of the pairwise probe functions dot-product instead of computing the singular value decomposition:

\begin{align}
    A_{ij} &= \sum_{\boldsymbol{r}} \mathcal{P}^{(i)*}(\boldsymbol{r}) \mathcal{P}^{(j)}(\boldsymbol{r}) = U \Lambda U^\dagger \label{eq:pairwise-dot-product-decomp} \\
    \mathcal{P}^{(i)'}(\boldsymbol{r}) &= \sum_j U_{ij} \mathcal{P}^{(j)}(\boldsymbol{r}). \label{eq:probe-orthogonalization}
\end{align}

\Cref{fig:mixed-state}f-g demonstrates the utility of the mixed-state formalism on the same STO [110] dataset reconstructed using a mixture of two orthogonal probe functions.
The algorithm successfully manages to separate the circular and annular apertures (\cref{fig:mixed-state}g) in the probe, which also results in a much sharper reconstructed object (\cref{fig:mixed-state}f) as compared to \cref{fig:mixed-state}d.
Finally, as alluded to earlier, the ability of the algorithm to successfully partition and orthogonalize the probe intensities strongly depends on the batch size $\mathcal{M}$.
Using a smaller batch size, such that (on-average) probe positions do not overlap substantially, ensures the algorithm performs more repeated adjoint and orthogonalization steps leading to more accurate probe intensities, albeit at an increased computational cost.

\begin{figure*}
\includegraphics[width=\textwidth]{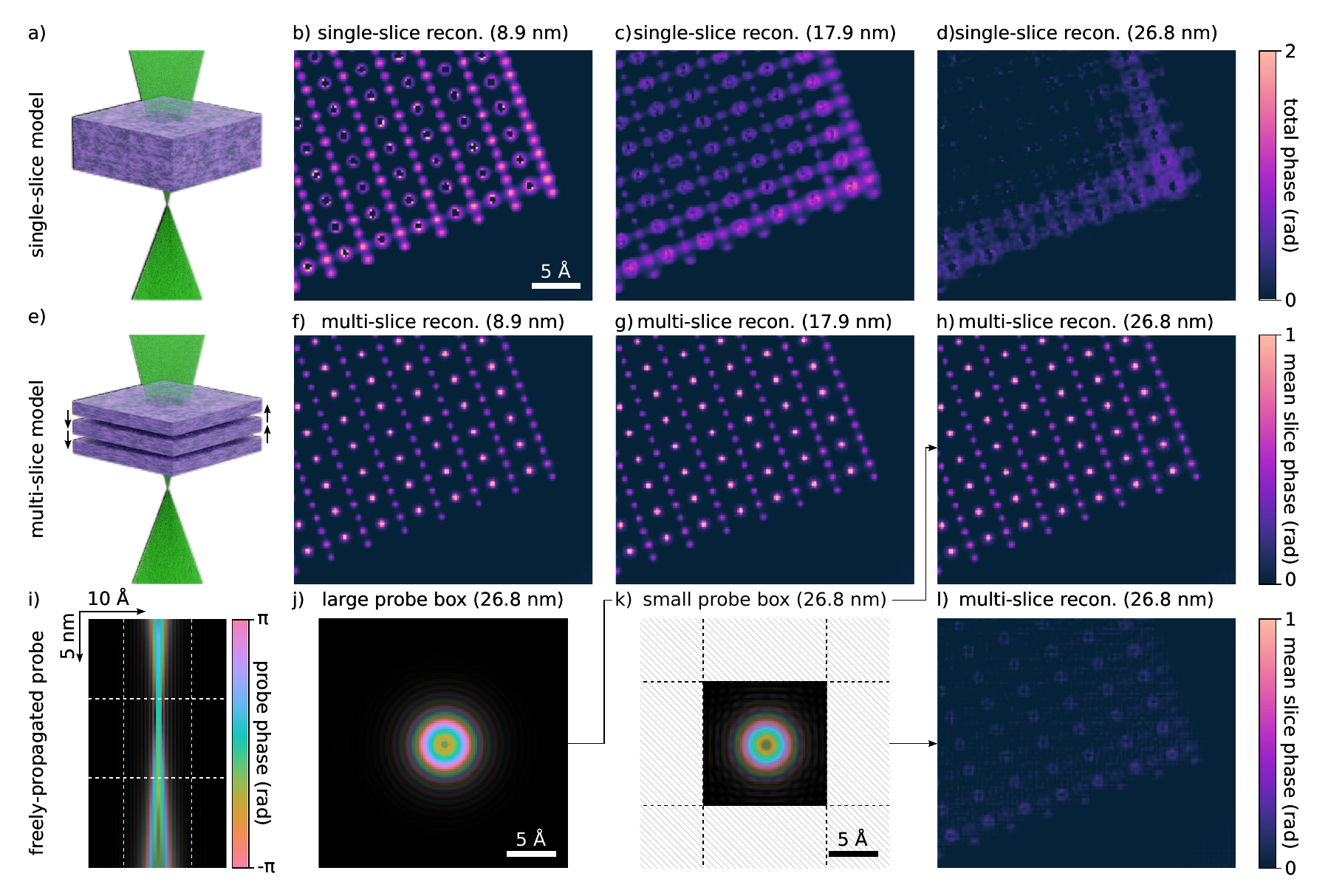}
\caption{
a-d) Single-slice and e-h) multi-slice reconstructions of simulated STO [110] slabs of various thicknesses, highlighting the breakdown of single-slice ptychography for ``thick'' strongly-scattering samples and the linearity of multi-slice ptychography. 
i) The probe changes size in real space as it is propagated.
j) A large cropping box is needed to hold the probe in real space at the exit surface for the 26.8 nm sample to k) avoid phase wrapping artifacts from a smaller box that l) cause artifacts in the sample reconstruction. 
}
\label{fig:multi-slice}
\end{figure*}

\section{Multi-slice ptychography} \label{sec:multi-slice_ptycho}

The forward models introduced so far in~\cref{sec:single-slice_ptycho,sec:mixed-state_ptycho} assume the exit-waves are given by a simple multiplication of the incoming-wave and the object.
While this is a convenient approximation for weakly-scattering or thin samples, it breaks down for multiple-scattering or thick samples as the illumination profile changes significantly as a function of depth due to dynamical scattering and free-space propagation~\citep{gao2017electron, clark2023effect}. 
To account for these effects, multi-slice ptychography approximates the object as a stack of $N$ slices~\citep{maiden2012ptychographic}, each thin-enough to satisfy the multiplicative assumption (\cref{eq:overlap-projection}), and modifies the forward operator to apply alternating transmission and propagation steps:

\begin{align}
    \psi_m^{(n)}(\boldsymbol{r}) &= \Pi_O\left[\phi_m^{(n)}, \mathcal{O}^{(n)} \right] = \phi_m^{(n)}(\boldsymbol{r}) \mathcal{O}^{(n)}(\boldsymbol{r}-\boldsymbol{R}_m),
    \intertext{where}
    \phi_m^{(1)}(\boldsymbol{r}) &= \mathcal{P}(\boldsymbol{r}) \\
    \phi_m^{(n)}(\boldsymbol{r}) &= \mathrm{Prop}_{\Delta z_{n-1}}\left[ \psi_m^{(n-1)}(\boldsymbol{r}) \right].
\end{align}
$\phi_m^{(n)}(\boldsymbol{r})$ and $\mathcal{O}_m^{(n)}(\boldsymbol{r}-\boldsymbol{R}_m)$ are the incoming-wave illumination and sample transmission functions evaluated at slice $n$, $\Delta z_n$ represents the thickness between slices $n$ and $n+1$, and the free-space propagator is defined according to:

\begin{align}
    \mathrm{Prop}_{\Delta z}\left[\cdot \right] &= \mathcal{F}^{-1} \left[ G_{\Delta z}(\boldsymbol{q}) \mathcal{F} \left[ \cdot \right] \right] \\
    G_{\Delta z}(\boldsymbol{q}) &= \mathrm{exp}\left(-i \pi \lambda \left| q \right|^2 \Delta z \right),
\end{align}
where $\lambda$ is the electron wavelength.
Note that for $N$ slices, we perform $N$ transmission steps but only $N-1$ propagation steps by convention to recover the single-slice formalism for $N=1$.

Starting from the last exit-wave ($n=N$) and working backwards, the adjoint operator is similarly modified to read, e.g. for the stochastic gradient descent algorithm:

\begin{align}
    \mathcal{O}^{(n)'}(\boldsymbol{r}) &= \mathcal{O}^{(n)}(\boldsymbol{r}) + \ddfrac{\beta \sum_{m \in \mathcal{M}} \phi_m^{(n)*}(\boldsymbol{r})\Delta \psi_m^{(n)}(\boldsymbol{r})}{|| \phi_m^{(n)}(\boldsymbol{r})||_{\alpha}} \label{eq:gd-update-o-multislice} \\
    \mathcal{P}'(\boldsymbol{r}) &= \mathcal{P}(\boldsymbol{r}) +\ddfrac{\beta \sum_{m \in \mathcal{M}} \mathcal{O}^{(1)*}(\boldsymbol{r}-\boldsymbol{R}_m)\Delta \psi_m^{(1)}(\boldsymbol{r})}{|| \mathcal{O}^{(1)}(\boldsymbol{r}-\boldsymbol{R}_m)||_{\alpha}} , \label{eq:gd-update-p-multislice}
    \intertext{where}
    \Delta \psi_m^{(N)}(\boldsymbol{r}) &= \Pi_F\left[\psi_m^{(N)}(\boldsymbol{r})\right]-\psi_m^{(N)}(\boldsymbol{r}) \label{eq:gd-psi-multislice-last} \\
    \Delta \psi_m^{(n)}(\boldsymbol{r}) &= \mathrm{Prop}_{-\Delta z_n} \left[ \Delta \psi_m^{(n+1)}(\boldsymbol{r}) \mathcal{O}^{(n+1)*}(\boldsymbol{r} -\boldsymbol{R}_m) \right]\label{eq:gd-psi-multislice-rest}. 
\end{align}

Note that this is fundamentally different than both the 3D e-PIE~\citep{maiden2012ptychographic} and maximum-likelihood~\citep{tsai2016xray} formalisms, as we do not take gradient steps at each slice, but rather back-propagate the single gradient step evaluated at the detector plane.

\Cref{fig:multi-slice} highlights the advantage of multi-slice ptychography for thick samples when the multiplicative assumption breaks down.  
In~\cref{fig:multi-slice}a-d, we examine the same field of view as~\cref{fig:mixed-state}a at various thicknesses using the single slice algorithm introduced in~\cref{sec:single-slice_ptycho}. 
Already at 8.9 nm the multiplicative assumption breaks down, which can be seen from the doughnut-like appearance of the strongly scattering strontium columns, indicative of phase wrapping. 
The reconstructions for thicker samples are further degraded (\cref{fig:multi-slice}c-d). 
By contrast, the multi-slice algorithm introduced above is able to recover the correct sample potential (\cref{fig:multi-slice}e-h), for thicknesses as large as 26.9 nm.
Specifically, here we use 1.1 nm thick slices for each reconstruction with strong regularization along the beam direction, requiring all slices to have equal phase shifts. 
Finally, we note that the mean slice phase for the various thicknesses in \cref{fig:multi-slice}f-h are plotted  on the same color scale, highlighting the linearity of the reconstruction. 

Of note are the additional defocus and region-of-interest probe box requirements for multi-slice ptychography.
As shown in~\cref{fig:multi-slice}i, as the probe is propagated in free-space, the size of the probe in real space changes. 
Moreover, this is an underestimate since the probe will further expand towards the exit surface thanks to scattering from beam-specimen interactions. 
As such, larger and larger region-of-interest cropping boxes are required to fit the probe with increasing thickness.
The multi-slice reconstructions in~\cref{fig:multi-slice} are preformed with the cropping shown in~\cref{fig:multi-slice}j, with the exit probe shown at largest thickness (26.8nm). 
Although a smaller cropping box (\cref{fig:multi-slice}k) may look adequate to hold the probe, some subtle wrapping artifacts are apparent for the exit wave from the thickest reconstructions.
This small change severely impacts the quality of the thickest reconstruction (\cref{fig:multi-slice}l).
For the thinner specimens (8.9 nm and 17.9 nm), there is less scattering and the probe travels a shorter distance, so the reconstructions with the smaller cropping box show no significant artifacts, with reconstructions nearly-identical to those in~\cref{fig:multi-slice}f-g.

Recently, multi-slice electron ptychography has been used to both achieve sub 20pm atomic-resolution limited by thermal vibrations~\citep{chen2021electron}, as well as demonstrate nanometer depth-resolution of dopant and vacancy concentrations and dislocations ~\citep{chen2021electron,yoon2023,ribet2024uncovering}. 
Proper regularization along the beam direction is essential for multislice reconstructions, and is discussed further in~\cref{subsec:object-regularization}.
In addition, high doses are required to achieve sufficient signal-to-noise to recover depth-dependent information. 

\begin{figure*}
\includegraphics[width=\textwidth]{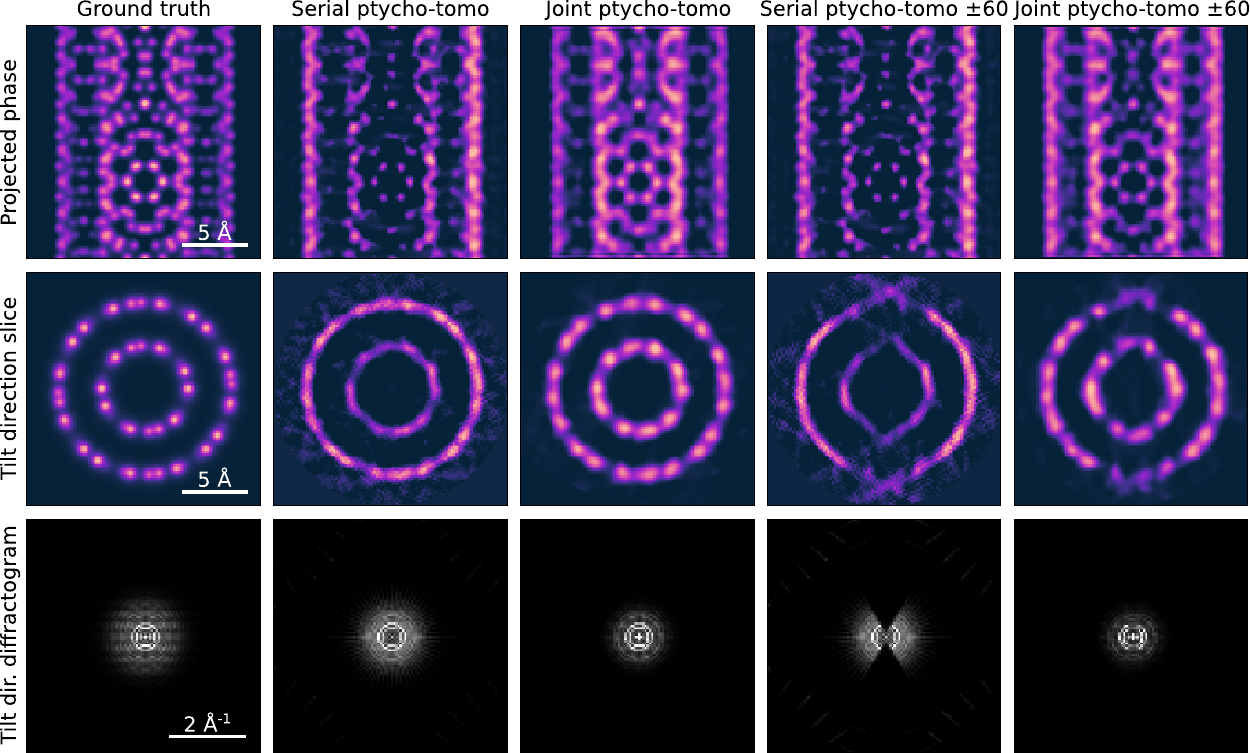}
\caption{
Serial and joint ptychographic tomography reconstructions of simulated double walled CNT, highlighting the importance of missing-wedge artifacts in serial ptychographic tomography, in contrast with the effective regularization of the missing-wedge in joint ptychographic tomography.
}
\label{fig:overlap-tomo}
\end{figure*}

\section{Joint Ptychographic Tomography} \label{sec:overlap-tomo}

Multi-slice ptychography provides limited resolution along the beam direction~\cite{chen2021electron}.
An alternative approach, with a rich literature in electron microscopy, is tilting the sample to obtain multiple two-dimensional projections and reconstructing them tomographically to obtain three-dimensional sample information~\citep{yang2017deciphering}.

While traditional STEM electron tomography techniques most often use the thermally-scattered electrons indicent on a high-angle annular dark-field (HAADF) detector, recent work has successfully reconstructed the three-dimensional atomic structure of a heterogeneous sample containing light elements using electron ptychography~\citep{pelz2023solving}.
In particular, the authors collected a tilt-series of 4D-STEM datasets of a double-walled CNT containing Zr-Te nanostructures and performed mixed-state ptychographic reconstructions for each tilt-angle, which were then reconstructed tomographically to obtain a three-dimensional volume. 
We will refer to this technique as \textit{serial ptychographic tomography}, due to the sequential fashion in which the analysis was performed.
\Cref{fig:overlap-tomo} illustrates this technique for a simulated tilt-series along the long-axis of the simulated CNT model shown in~\cref{fig:dpc}b with six-degree increments.
While the full tilt-series reconstruction (\cref{fig:overlap-tomo}, second column) shows sufficient reconstruction quality to trace atoms, the missing wedge artifacts shown in the fourth column of~\cref{fig:overlap-tomo} severely degrade the reconstruction quality.
In addition to being much more dose-efficient compared to HAADF tomography, serial ptychographic tomography can take advantage of the aforementioned benefits of various flavours of ptychography to reconstruct thicker samples with partial coherence.
However, it still suffers from common tomographic artifacts such as the missing-wedge.

An alternative, even more dose-efficient technique is to perform \textit{joint ptychographic tomography} whereby the three-dimensional volume is reconstructed directly from the 4D-STEM tilt-series dataset~\citep{lee2023}.
The recently-proposed technique, also referred to as \textit{multi-slice electron tomography}, extends multi-slice ptychography by successively rotating the three-dimensional object estimate, such that the beam direction is aligned with the z-axis for each tilt, and updating the intensity along z- ``super'' slices using~\cref{eq:gd-update-o-multislice,eq:gd-update-p-multislice,eq:gd-psi-multislice-last,eq:gd-psi-multislice-rest}.
The advantages of jointly updating the three-dimensional volume directly are shown in the third and fifth columns of~\cref{fig:overlap-tomo} for the full tilt-series and 60$^{\circ}$ missing-wedge series respectively.
In particular, the additional angular redundancy in the dataset, combined with effective regularization~\cref{sec:experimental-considerations}, enable the reconstruction to efficiently fill in information along the missing-wedge direction (\cref{fig:overlap-tomo}, bottom-right panel), resulting in an overall superior reconstruction.

\section{Experimental Considerations} \label{sec:experimental-considerations}

The non-convexity and high-dimensionality of the ptychographic inverse problems described in~\cref{sec:single-slice_ptycho,sec:mixed-state_ptycho,sec:multi-slice_ptycho,sec:overlap-tomo} make them particularly hard to converge, especially in the presence of experimental noise.
Additionally, the algorithms are not guaranteed to converge to optimal or unique solutions.
This can be in-part remedied using effective regularization, which reduces the dimensionality of the available solution-space or constrains ill-posed optimization problems.

While some of these regularizations can be implemented by adding additional priors or penalty terms in the loss function (\cref{eq:nmse-error}) directly, this can often result in increased computational complexity for the gradient step.
Instead, we implement a projected gradient approach, whereby we first take unconstrained gradient steps using for example~\cref{eq:gd-psi,eq:gd-update-o,eq:gd-update-o,eq:proj-psi,eq:proj-update-o,eq:proj-update-p} and then, at every iteration, project the current probe and object estimates to feasible sets subject to the constraints described below.
This can be significantly more efficient computationally, as long as the constraint projections are computationally inexpensive to evaluate at every iteration.
In general, as little regularization as possible should be applied to avoid spurious artifacts in the probe and object reconstructions.

\subsection{Object regularization} \label{subsec:object-regularization}

Object regularization includes ensuring the reconstructed object is smooth, using Gaussian or Butterworth filters in real- and reciprocal-space respectively, as well as total variation denoising. 
These smoothing filters can be applied both in the plane of the object and along the beam-direction for multi-slice and joint ptychographic tomography. 
For accurate recovery of slices, multi-slice ptychography in particular requires regularization along the beam-direction~\citep{chen2021electron}. 
The strongest such regularization is that all slices have identical phase contributions. 
This strict assumption can be relaxed by instead enforcing first-order total variation denoising, allowing for non-identical slices which ideally approach scattering from the substrate/vacuum at large distances from the sample.

Separating the phase and amplitude components of the complex-valued reconstructed object is a challenge in electron ptychography.
For weakly-scattering objects (or equivalently sufficiently thin slices of strongly-scattering objects in multi-slice ptychography), the amplitude of the reconstructed object should vary negligibly from unity.
As such, we can reduce the solution-space dimensionality by a factor of two by employing the pure-phase approximation and constraining the amplitude to unity at each iteration.
Taking this a step further, the object itself can be stored as the real-valued potential directly, $\mathcal{O}(\boldsymbol{r}) = \mathrm{e}^{\mathrm{i} V(\boldsymbol{r})}$.
In addition to the memory efficiencies this affords, it further allows us to impose additional regularization constraints on the potential directly.
Notably, atomic potentials are, by nature, non-negative and thus we can enforce positivity by clipping negative values to zero.
Moreover, positivity can be combined with shrinkage filters, whereby a constant phase-shift is subtracted from the current estimate at each iteration prior to zero-clipping, to promote atomicity by forcing the background to zero.
We note however that this constraint is rather aggressive, especially in early iterations, and can result in artificially suppressing scattered signal.

\begin{figure*}
    \centering
    \includegraphics[width=\textwidth]{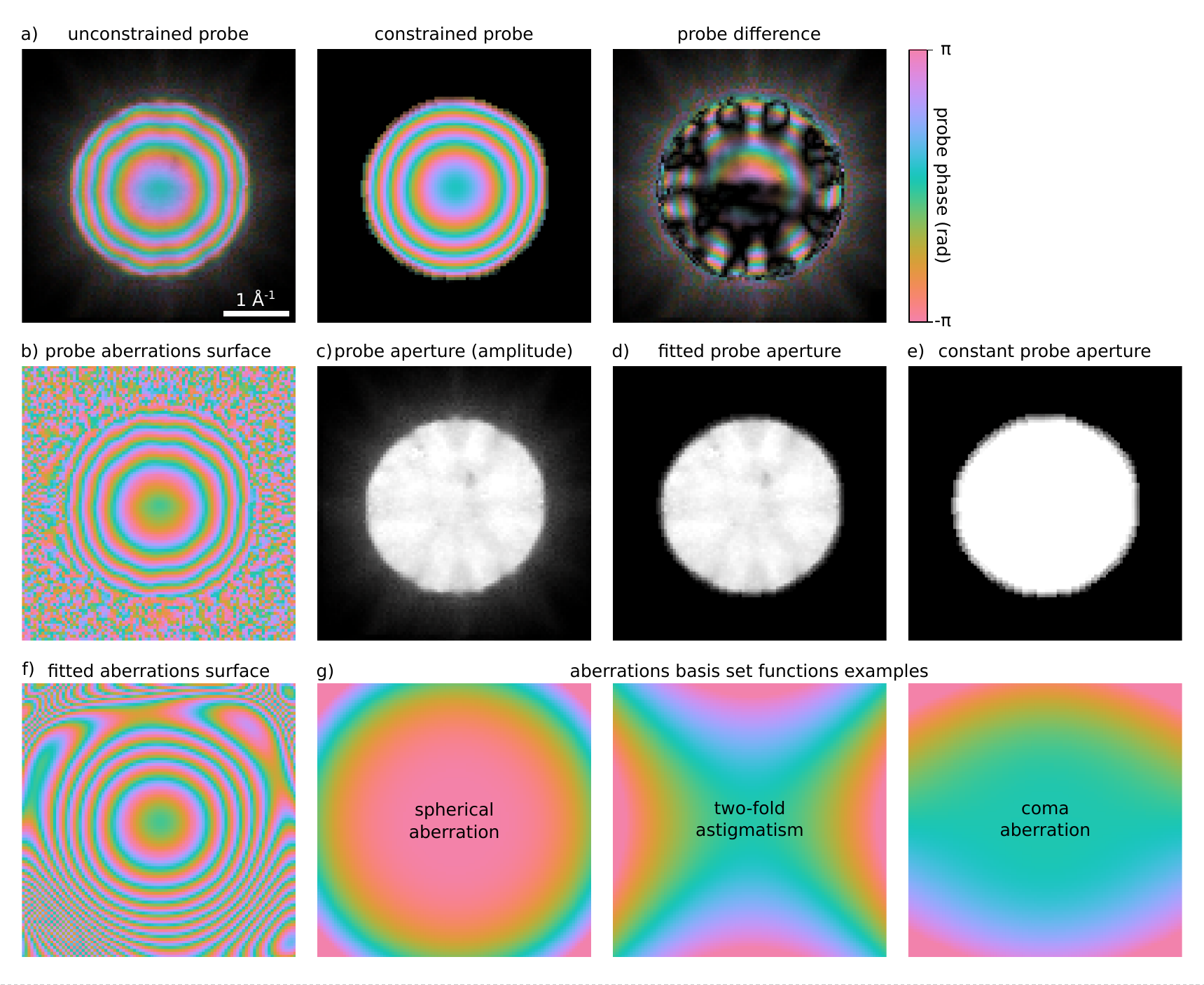}
    \caption{a) Unphysical reconstructed probe illustrating intensity ``leakage'' outside the probe aperture, probe-object mixing seen by the eight-fold symmetry features, and non-smooth aberration surface.
    The probe can be separated to b) aberration surface and c) aperture functions. 
    The probe aperture can be constrained to d) remove intensity outside the aperture and e) exhibit flat-field intensity inside the aperture.
    Similarly, the probe aberration surface, can be fitted to a linear combination of g) low-order aberrations to produce a f) smooth estimate of the surface.
    Panel a) also shows a comparison between the complex-valued unconstrained probe (left) and constrained probe (middle), together with their difference (right).
    }
    \label{fig:probe-regularization}
\end{figure*}

\subsection{Probe regularization} \label{subsec:probe-regularization}

Similarly, the current probe estimate can be constrained using physical regularizations, such as a fixed probe aperture, a smooth phase aberration surface, and center-of-mass fixing.
As mentioned in~\cref{subsec:ptycho-preprocessing}, the probe aperture can be initialized using either a vacuum measurement or an estimate of the probe semi-angle, i.e. a perfect circular aperture.
In general, a high signal-to-noise vacuum measurement of the probe aperture is important for probe regularization and calibration. 
This can be acquired by averaging several diffraction intensity measurements over a vacuum region with the same experimental parameters as the 4D-STEM dataset, albeit with a smaller step-size to avoid descan artifacts.

A common artifact in ptychographic reconstructions is probe-object mixing (\cref{fig:probe-regularization}a), where object features from the scattered diffraction intensities appear in the Fourier amplitude of the probe. 
This can be ameliorated by replacing the Fourier amplitude of the probe estimate using the square-root of a vacuum measurement, effectively reducing the solution-space dimensionality by a factor of two.
If such a measurement is not available (and cannot be estimated from a vacuum region inside the field of view), then the Fourier amplitude of the probe estimate can be fitted at every iteration to an angularly-varying sigmoid function to suppress intensity outside the probe aperture and ensure a constant top-hat intensity inside the aperture (\cref{fig:probe-regularization}c-e).

Moreover, the probe aberration surface is expected to be smoothly-varying.
This is not guaranteed \textit{a-priori} (\cref{fig:probe-regularization}b), and can instead be imposed as an iterative constraint.
An elegant way of enforcing this is to fit the aberration surface $\chi(\boldsymbol{k})$ to a linear combination of low-order aberrations~\citep{ophus2016aberrations}, by restricting the maximum radial and angular orders of the expansion in~\cref{eq:chi-expansion}.
\Cref{fig:probe-regularization}f-g show the fitted aberration surface and some basis set examples respectively. 
Finally, we note that for this to work, the aberration surface needs to be phase-unwrapped first which can introduce additional sources of error.
If an accurate initial estimate of the probe aberrations exists, e.g. using the parallax aberration coefficients fitting introduced in~\cref{subsec:parallax-preprocessing-results}, these can be removed prior to smoothing and re-introduced after to avoid having to phase-unwrap the aberration surface.

\subsection{Probe positions updates} \label{subsec:positions-regularization}

Because ptychographic algorithms rely on overlap between adjacent probe positions, accurate probe positions are required for object convergence~\citep{maiden2012annealing}.
The probe positions estimate can be refined during the ptychographic reconstruction as well as constrained to fit a linear drift distortion model. 
Drift in STEM data acquisition causes the real probe positions to deviate from the ideal raster positions~\citep{ophus2016correcting}.
In general, faster 4D-STEM data acquisition times, which are determined by detector read-out speeds, step size, and the required field of view, lead to fewer artifacts in the positions of the probes.
 
During ptychographic reconstruction, we can refine our probe positions estimate using a gradient step on the intensity measurements in parallel~\citep{dwivedi2018}.
This results in a displacement vector for each probe position, which can be further constrained by fitting the displacement vectors to a six-parameter (x-scaling, y-scaling, shear, rotation, x-translation, y-translation) global affine transformation. 
 
\subsection{Hyper-parameter Tuning} \label{subsec:parameter-tuning}

In order to successfully perform a ptychographic reconstruction of experimental data, it is necessary that certain experimental parameters be known to high precision, particularly the scan step size, the reciprocal pixel size of the diffraction patterns, and the real-space/reciprocal-space rotation angle. 
Other parameters, such as the defocus and other aberrations are refined during reconstruction, but a better guess still affects the convergence. 
In addition, reconstruction parameters such as the step size, batch size, and regularization strengths, also impact the quality of the reconstruction. 
In order to find the optimum values of both experimental and reconstruction parameters that give the highest quality reconstruction of a dataset, we have implemented an optimizer using the Bayesian optimization (BO) with Gaussian processes (GP) algorithm. 
BO-GP is commonly used in the tuning of hyperparameters for machine learning~\citep{bergstra_algorithms_2011}, and has been successfully applied to ptychography both for parameter tuning~\citep{cao_automatic_2022} and directly as a procedure for iterative phase retrieval~\citep{pelz_low-dose_2017}. 
The BO-GP algorithm models the function to be minimized as a mixture of Gaussian processes, and chooses evaluation points through a combination of exploration of unsampled regions of the parameter space and exploitation of local minima of the GP model.
Our implementation provides a simple interface for specifying optimization parameters, and the optimization is performed by the BO-GP routine provided by \texttt{scikit-optimize}~\citep{head_scikit-optimizescikit-optimize_2021}.
Various reconstruction quality metrics can be used as the objective function of the optimizer, including the final data error of the reconstruction as well as the contrast, entropy, or total variation of the reconstructed object. 

\subsection{Data Acquisition Recommendations} \label{subsec:data-acquisition}

While the ideal experimental parameters depend strongly on the research question, the microscope, and the specimen, there are a few suggestions that may be helpful for most phase retrieval experiments. 
First, we recommend starting by simulating a proxy of the system to test the range of experimental parameters that may be possible with a given set-up. 
For example, the number of pixels in available detectors, combined with the convergence semi-angle limit the maximum defocus that is possible. 
When recording diffraction data, it is important the detector can sample the diffraction intensities finely enough to recover features of the specimen in the ronchigram, which further set constraints on the maximum possible defocus value. 

In designing an experiment, it is also important to consider realistic limitations of the microscope and specimen. 
For example, although the methods introduced above can provide some degree of probe-position refinement, the reconstruction will be severely limited with substantial scan distortion.
Therefore, the number of probe positions in a scan should be determined based on stability, magnification, and detector speed.
Similarly, while ptychography and parallax can correct for aberrations in the incident beam, when possible it is better to align the microscope as well as possible before starting the experiment, to remove higher-order residual aberrations. 
There can be other sources of artifacts in experimental data that are not usually accounted for ptychographic models, such as incoherence of the source, post-specimen aberrations introduced by the beam through a spectrometer, and dead detector pixels.
All of these should be minimized when possible.
This may require using a smaller beam current to minimize incoherence, aligning the spectrometer, dark references, and preprocessing the diffraction data. 
Preprocessing may include weighted-binning to account for dead pixels, hot- and cold-pixel filtering, and gain correction.

Further, calibration scans can be helpful for initializing parameters for phase retrieval experiments. 
These scans may include experiments on samples, such as the gold nanoparticles shown in~\cref{fig:gold},
which provide a large phase shift, used to calibrate the rotation between real and reciprocal space.
As described above, it is important to use the right range of defocus values for ptychography experiments. 
However, the nominal microscope defocus value and true defocus value (and sign) might not agree. 
Therefore, acquiring calibration scans at a few defocus values can solve for the relationship between indicated and true defocus. 
Calibrations may be similar day-to-day on the same microscope especially for similar alignment conditions. 
Therefore, these measurements are most important for initial phase retrieval experiments on a microscope. 

As highlighted a few times in this manuscript, vacuum scans are invaluable for ptychography experiments. 
For initializing the probe for ptychography and possibly constraining it during the reconstruction, a vacuum scan should produce a high signal-to-noise image of the aperture. 
Ideally this is done with the same diffraction shift, so the vacuum and data are centered on the detector. 
A vacuum scan does not requre as many images as a full 4D-STEM scan, since all the scan positions are averaged together to create one 2D vacuum probe.
Instead it may be better to take fewer scan positions at a higher dose, if allowed by the detector. 
Although descan can be accounted for computationally after the experiment, it is better to minimize this artifact in vacuum measurements by taking the scan at as high magnification as possible, or keeping the STEM probe stationary while recording these images.
It may also be helpful to take a vacuum scan with a very large camera length, so the probe is evenly illuminating the entire detector. 
This flat-field reference can be used for gain correction of the 4D-data.
Lastly, when encountering any unexpected challenges on the microscope, it is always a good idea to consider recruiting a few additional rubber ducks to help you debug the issue.

\begin{table}
\begin{tabular}{l r c l}
\toprule
 & Symbol & & Definition \\
 \cmidrule(lr){2-2}
 \cmidrule(lr){4-4}
& & & \\

\multicolumn{2}{l}{General} & & \\
\cmidrule(lr){1-2} 
 & $\mathcal{F}\left[ \cdot \right]$ & $=$ & 2D Fourier transform operator \\
 & $\Re\left\{ \cdot \right\}$ & $=$ & Real part of complex expression \\
 & $\sign\left\{ \cdot \right\}$ & $=$ & Sign of real expression \\
 & $\nabla$ & $=$ & Spatial derivative operator \\
 & $\boldsymbol{r}$ & $=$ & Real-space coordinate \\
 & $\boldsymbol{k}$ & $=$ & Reciprocal-space coordinate \\
 & $\boldsymbol{R}_m$ & $=$ & m\textsuperscript{th} real-space probe position \\
 & $I(\boldsymbol{R}_m,\boldsymbol{k})$ & $=$ & 4D-STEM diffraction intensities \\
 & $A(\boldsymbol{k})$ & $=$ & Fourier-space probe aperture \\
 & $\chi(\boldsymbol{k})$ & $=$ & Fourier-space aberration function \\
 & $\beta$ & $=$ & Step size \\
 & $\mathcal{E}$ & $=$ & Error metric \\
 & & & \\

\multicolumn{2}{l}{Iterative DPC} & & \\
\cmidrule(lr){1-2}

& $V(\boldsymbol{r})$ & $=$ & Electrostatic sample potential \\
& $\mathcal{I}_{\alpha}^V(\boldsymbol{R})$ & $=$ & Virtual CoM image along direction $\alpha$\\
& & & \\

\multicolumn{2}{l}{Parallax} & & \\
\cmidrule(lr){1-2}
& $\boldsymbol{w}(\boldsymbol{k})$ & $=$ & virtual image shifts  \\
& $\boldsymbol{v}(\boldsymbol{k})$ & $=$ & position on bright-field disk  \\
& $C_{m,n}^x$ \& $C_{m,n}^y$ & $=$ & Orthogonal aberration coefficients of \\
& & & radial and angular order $(m+1, n)$ \\
& $C_{1,0}$ & $=$ & (Negative) defocus \\
& $C_{1,2}^x$ \& $C_{1,2}^y$& $=$ & Astigmatism  \\
& & & \\

\multicolumn{2}{l}{Ptychography} & & \\
\cmidrule(lr){1-2}
& $\mathcal{P}(\boldsymbol{r})$ & $=$ & Real-space probe operator \\
& $\mathcal{O}(\boldsymbol{r}-\boldsymbol{R}_m)$ & $=$ & Real-space object operator \\
& $\Pi_O\left[ \cdot \right]$ & $=$ & Overlap projection operator \\
& $\Pi_F\left[ \cdot \right]$ & $=$ & Fourier projection operator \\
& $J_m(\boldsymbol{k})$ & $=$ & Estimated model intensities \\
& $\psi_m(\boldsymbol{r})$ & $=$ & Real-space exit-waves \\
& $\Delta \psi_m^{\mathrm{GD/PG}}(\boldsymbol{r})$ & $=$ & Real-space exit-waves gradient \\
& $||\cdot||_{\alpha}$ & $=$ & $\alpha$-regularized norm \\
& $(a,b,c)$ & $=$ & Projection-set parameters \\
& & & \\

\multicolumn{2}{l}{Info. Transfer} & & \\
\cmidrule(lr){1-2}
& $ \Delta \phi^{\mathrm{syst}}(\boldsymbol{k})$ & $=$ & Systematic phase error \\
& $ \Delta \phi^{\mathrm{rand}}(\boldsymbol{k})$ & $=$ & Random phase error \\
& $ \Delta \phi^{\mathrm{total}}(\boldsymbol{k})$ & $=$ & Total phase error \\
& & & \\

\bottomrule
\end{tabular}
\caption{
Nomenclature used in the manuscript.
For mixed-state, multi-slice, and joint ptychographic tomography, we try and use consistent notation with the appropriate superscript changes to denote the probe, slice, and tilt indices respectively.}
\label{table:1}
\end{table}

\end{appendices}

\section{Availability of Data and Materials}
The simulated and experimental datasets used in this study are available at [\hl{add zenodo link}].
The simulation and analysis notebooks used in this study are available at \href{https://github.com/gvarnavi/py4dstem-phase-retrieval-paper-notebooks}{https://github.com/gvarnavi/py4dstem-phase-retrieval-paper -notebooks}.
The open-source software package used in this study is available at \href{https://github.com/py4dstem/py4DSTEM}{https://github.com/py4dstem/py4DSTEM}.

\section{Author Contributions}
GV, SMR, and CO conceived of the study, implemented the core phase retrieval algorithms, analyzed the simulated and experimental data, and wrote the initial draft of the manuscript.
SEZ implemented the Bayesian optimization framework.
SMR acquired the experimental datasets, together with DOB and FIA (hBN dataset) and YY (VLPs dataset).
BHS implemented the initial framework for the open-source software this study extends.
VPD, MCS, and CO acquired funding for the study.
All authors contributed to the review and editing of the manuscript.

\section{Acknowledgments}
We thank Lena Kourkoutis for her inspirational ideas and support of this project.
We also gratefully acknowledge fruitful discussions with Joel E. Moore of the University of California, Berkeley.

\section{Financial Support}
GV acknowledges support from the Miller Institute for Basic Research in Science.
SMR acknowledges support from the IIN Ryan Fellowship and the 3M Northwestern Graduate Research Fellowship, and the U.S. Department of Energy, Office of Science, Office of Workforce Development for Teachers and Scientists, Office of Science Graduate Student Research (SCGSR) program. 
The SCGSR program is administered by the Oak Ridge Institute for Science and Education for the DOE under contract number DE‐SC0014664.
BHS acknowledges support from the Toyota Research Foundation.
CO and SMR acknowledge support from the US Department of Energy Early Career Research Program. 
DOB acknowledges support from the Department of Defense through the National Defense Science \& Engineering Graduate (NDSEG) Fellowship program.
This work made use of the electron microscopy facility of the Platform for the Accelerated Realization, Analysis, and Discovery of Interface Materials (PARADIM), which is supported by the National Science Foundation under Cooperative Agreement No. DMR-2039380.
This work received support from the SHyNE Resource (NSF ECCS-2025633).
Work at the Molecular Foundry was supported by the Office of Science, Office of Basic Energy Sciences, of the U.S. Department of Energy under contract number DE-AC02-05CH11231.

\section{Competing Interests}
The authors declare that they have no competing interests.

\end{document}